\documentclass[conference]{IEEEtran}
\IEEEoverridecommandlockouts
\let\labelindent\relax
\usepackage{enumitem}
\usepackage{ulem}
\usepackage{nicefrac}
\usepackage{siunitx}
\usepackage{array,framed}
\usepackage{amsthm}
\usepackage{booktabs}
\usepackage{
  color,
  float,
  epsfig,
  wrapfig,
  graphics,
  graphicx,
  subcaption
}
\usepackage{etoolbox}
\usepackage{textcomp,amssymb}
\usepackage{setspace}
\usepackage{latexsym,fancyhdr,url}
\usepackage{enumerate}
\usepackage{algorithm}
\usepackage[noend]{algpseudocode}
\usepackage{graphics}
\usepackage{xparse} 
\usepackage{xspace}
\usepackage{multirow}
\usepackage{csvsimple}
\usepackage{balance}

\usepackage{
  tikz,
  pgfplots,
  pgfplotstable
}
\usepackage{hyperref}

\usetikzlibrary{
  shapes.geometric,
  arrows,
  external,
  pgfplots.groupplots,
  matrix
}
\pgfplotsset{compat=1.18} 

\theoremstyle{plain}

\theoremstyle{definition}

\usepackage{mathtools}
\usepackage{cite}
\usepackage{amsmath,amssymb,amsfonts}
\usepackage{graphicx}
\usepackage{textcomp}
\usepackage{xcolor}
\def\BibTeX{{\rm B\kern-.05em{\sc i\kern-.025em b}\kern-.08em
    T\kern-.1667em\lower.7ex\hbox{E}\kern-.125emX}}


\begin{document}

\title{Dagger: Decoupling-based Model Stealing Attack against Graph Neural Networks}

\author{\IEEEauthorblockN{Anonymous Author(s)}}





\author{\IEEEauthorblockN{Ying Song}\IEEEauthorblockA{\textit{University of Pittsburgh}\\yis121@pitt.edu}\and\IEEEauthorblockN{Xiaowei Jia}\IEEEauthorblockA{\textit{Rutgers University}\\xiaowei.jia@rutgers.edu}\and\IEEEauthorblockN{Balaji Palanisamy}\IEEEauthorblockA{\textit{University of Pittsburgh}\\bpalan@pitt.edu}}


\maketitle

\begin{abstract}

As Graph Neural Networks (GNNs) are widely deployed as Machine Learning-as-a-Service (MLaaS) APIs, model stealing attacks have emerged as a critical security threat. By querying a victim model's black-box API, an adversary can construct a functionally equivalent surrogate model, compromising proprietary intellectual property and downstream security. Existing GNN stealing attacks, however, rely on overly permissive assumptions, such as soft-label outputs, large query budgets, full-graph query access, and prior knowledge of victim backbones that rarely hold in real-world deployments. In this work, we formalize a strictly constrained black-box, hard-label and backbone-agnostic threat model for GNN stealing attacks under a tight query budget. Given these realistic restrictions, we identify four fundamental challenges: sparse local structures and isolated nodes that degrade victim label quality, insufficient supervision signals, systematic imbalance with incomplete class coverage, and backbone mismatch. To address these interlocking barriers, we propose Dagger, a novel two-phase decoupling-based attack framework. Specifically, in Phase 1, Dagger pre-trains a surrogate using decoupled information propagation to preserve structural context over sparse local subgraphs while handling isolated nodes, combined with manifold-level node mixup to synthesize continuous supervision signals and smooth decision boundaries. In Phase 2, Dagger freezes the encoder and fine-tunes the classifier head via class-balanced sampling paired with logit adjustment to rectify severe query imbalance without requiring extra victim queries. Extensive experiments across four benchmark graphs and four GNN backbones demonstrate that Dagger consistently outperforms state-of-the-art GNN stealing attacks, achieving up to 18.16\% higher fidelity while only utilizing 12.23$\times$ fewer queries than the strongest baseline. Furthermore, Dagger consistently bypasses state-of-the-art query-monitoring and backdoor-based watermarking defenses, underscoring the urgent need for defense paradigms tailored to such realistic GNN stealing threats.


\end{abstract}


\section{Introduction}
\label{sec:intro} 


Graph Neural Networks (GNNs) have showcased remarkable performance across diverse real-world applications, ranging from drug discovery \cite{fang2025recentdevelopmentsgnnsdrug} and fraud detection \cite{fraud} to recommendation systems \cite{wu2022graph}. Since training proprietary GNNs demands expensive data curation, computational resources and domain expertise, model providers increasingly commercialize their models via black-box Machine Learning-as-a-Service (MLaaS) APIs. However, their economic value renders GNNs attractive targets for model stealing attacks, where adversaries issue queries to extract functionally equivalent surrogate models, compromising proprietary intellectual property. Moreover, such stolen GNNs can serve as a springboard for subsequent adversarial attacks, such as backdoor injection and privacy inference attacks, thereby threatening broader system security.


Despite a growing body of work on model stealing attacks against GNNs \cite{GNN_MSA_2019, GNN_MSA_2022_Asia_CCS, IGNN_MSA_2022_SP, IGNN_MSA_2024, RealisticGNN_MSA_2024_KBS, DF_GNN_MSA_2024_USENIX, wang2025cega}, existing methods suffer from five critical limitations that severely hinder their practical applicability. (1) Over-reliance on rich information leakage: prior works assume APIs return full posterior probability or node embeddings \cite{wang2025cega, IGNN_MSA_2022_SP, IGNN_MSA_2024}, whereas model providers often restrict outputs to hard labels due to strict security and privacy policies in reality. (2) Topological over-privilege: current hard-label attacks still perform queries implicitly on the full graph or multi-hop neighborhoods beyond legitimate access \cite{GNN_MSA_2019,GNN_MSA_2022_Asia_CCS,RealisticGNN_MSA_2024_KBS}, granting structural information that would be unavailable in a realistic black-box setting. (3) Large query budgets and detection risks: even data-free attacks that avoid using any real data still resort to synthetic data generation techniques, such as generative adversarial network (GAN)-based graph generation \cite{DF_GNN_MSA_2024_USENIX}. 
These approaches always incur high computational overhead, training instability, and statistically anomalous query patterns that can be readily detected by 
query-monitoring defenses \cite{prada}. (4) Victim backbone dependency: the vast majority of GNN stealing attacks heavily depend on specific victim architectures or assume prior knowledge of the target backbone, which rarely holds in real-world deployments. Under a true black-box setting, backbone misalignment may severely degrade surrogate performance due to divergent information propagation dynamics. (5) Evaluation inconsistency: while prior works claim to operate under restricted black-box threat models \cite{wang2025cega,GNN_MSA_2022_Asia_CCS}, their open-source implementations implicitly grant the adversary access to the full graph during inference. This privilege inflates reported attack performance, suggesting that the practical severity of GNN stealing attacks under strict black-box constraints has been systematically underestimated and motivating a rigorous re-examination under a more faithful threat model.

To uncover the true vulnerabilities of GNNs under realistic MLaaS deployments, we formalize a strictly constrained black-box, hard-label, and backbone-agnostic threat model for GNN stealing attacks under a tight query budget. Specifically, each query is strictly executed on the local subgraph induced by the adversary's query node set, preventing access to the global topological context. Under this threat model, we identify four fundamental challenges that existing methods fail to address.

\begin{itemize}
    \item \textbf{Sparse local structures and isolated nodes}: queries over induced subgraphs frequently collapse into sparse topologies or isolated nodes, particularly on sparse or small-scale graphs. Standard GNNs, which rely on multi-hop neighborhood aggregation, degrade significantly under such degenerative structures, producing unreliable victim predictions that corrupt the surrogate's training signals.
    \item \textbf{Insufficient supervision signals:} Discrete hard labels discard inter-class confidence distributions, leaving the surrogate with insufficient supervision to learn smooth decision boundaries under a strict query budget. Unreliable victim predictions further propagate label noise into surrogate training, severely distorting the surrogate's decision boundaries across class margins.   
    \item \textbf{Systematic imbalance and incomplete class coverage:} Structural dominance in sparse subgraphs yields systematic biases, where majority classes receive a disproportionate share of labels while the minority remain severely under-represented or even uncovered by victim predictions.
    \item \textbf{Victim Backbone Mismatch:} The misalignment between the unknown victim and surrogate backbones further amplifies the aforementioned issues.
\end{itemize}

To tackle these interlocking challenges, we propose a novel \underline{d}ecoupling-b\underline{a}sed model stealin\underline{g} attack against \underline{g}raph n\underline{e}u\underline{r}al networks (Dagger). This framework consists of two phases, namely \textit{Decoupling-based Surrogate Pre-training} and \textit{Imbalance-aware Classifier Tuning}. In the first phase, motivated by APPNP \cite{appnp} to decouple feature transformation from graph propagation, we pretrain a topology-adaptive surrogate using a decoupled architecture to preserve structural context over sparse local subgraphs while maintaining meaningful representations for isolated nodes. To compensate for hard-label coarseness and incomplete class coverage, we perform manifold-level node mixup, which interpolates hidden representations across class boundaries. In the second phase, to rectify severe query-induced class imbalance, we freeze the surrogate encoder and only fine-tune the classifier head through class-balanced sampling paired with logit adjustment \cite{longtail_la}, providing complementary calibrations at both gradient and prediction levels to improve surrogate fidelity on minority classes without requiring extra victim queries. 

We evaluate Dagger across four representative benchmark graph datasets and four victim GNN backbones, demonstrating both superior stealing effectiveness and query efficiency. Notably, with 12.23$\times$ fewer queries than the strongest baseline \cite{GNN_MSA_2022_Asia_CCS}, Dagger still achieves up to 18.16\% higher fidelity. With the same query budget, Dagger outperforms the rest baselines by up to 45.27\% in fidelity, particularly on sparse and class-imbalanced graphs. Furthermore, we assess Dagger against two state-of-the-art defense paradigms: a query-monitoring detector \cite{prada} and a backdoor-based watermarking mechanism \cite{backdoorwm}. Since Dagger issues queries via real local subgraphs, its query patterns exhibit no distribution-level anomalies. Additionally, the strict yet realistic constraints prevent watermark triggers from being transferred to the surrogate. The experimental validation confirms that Dagger consistently bypasses both defenses, underscoring the urgent need for defense mechanisms tailored to realistic GNN stealing threats.

We summarize our contributions as follows:
\begin{itemize}
\item \textbf{Realistic Threat Formalization}: We systematically uncover five critical bottlenecks of existing stealing methods and formalize a strictly constrained black-box, hard-label, and backbone-agnostic threat model for GNN stealing attacks under a tight query budget. 
\item \textbf{Novel Decoupled Framework}: We propose Dagger, a two-phase decoupling-based attack framework that leverages decoupled information propagation to handle structural sparsity, applies manifold-level node mixup to enhance representation diversity and class coverage, and employs class-balanced sampling paired with logit adjustment to resolve severe query imbalance.
\item \textbf{Extensive Empirical Validation}: We conduct comprehensive evaluations across diverse graph datasets, victim backbones, and defense mechanisms. The empirical results confirm Dagger's superior effectiveness and efficiency, while demonstrating its resistance to both query-monitoring and watermarking defenses.

\end{itemize}

\section{Background and Related Work}
\label{sec:background}

\subsection{Node Classification}
Given an undirected attributed graph $\mathcal{G}\!=\!(\mathcal{V}, \mathcal{E}, X)$, $\mathcal{V}$ denotes a node set with $|\mathcal{V}|$ nodes and each node $v$ is associated with a feature vector $X_v\in\mathcal{R}^{1\times d}$, where $d$ is the dimension of node features, $\mathcal{E}$ represents an edge set with $|\mathcal{E}|$ edges, GNNs $\Phi(\mathcal{G})$ aggregate each node $v\in\mathcal{V}$'s information from its local neighborhood $\mathcal{N}(v)$ and further update its node embedding $H^{l}_{v}$ at the $l$-th layer. Formally, this process can be expressed as:
\begin{equation}
    H_{v}^{l}=UPD^{l}(H_{v}^{l-1},AGG^{l-1}(\{H_{u}^{l-1}:u\in\mathcal{N}(v)\}))
\end{equation}
where $H_{v}^{0}\!=\!X_v$, $l \in \{1,\ldots,L\}$ 
with $L$ denoting the total number of layers. $UPD$ and $AGG$ are two arbitrary differentiable functions to design diverse GNN backbones. For node classification tasks, generally, $H_{v}^{L}$ is fed into a linear classifier $f$ with a softmax function to obtain the final prediction $\hat{Y}$.

\subsection{Model Stealing Attacks against GNNs}

Existing model stealing attacks against GNNs can be broadly divided into two categories: query-based and data-free attacks. Query-based attacks assume the adversary can interactively query the victim GNN with partial or full publicly available graphs to receive model responses, while data-free attacks restrict the adversary's access to any real data, she/he relies on sophisticated data synthesis and  optimization techniques to ceaselessly query the victim with crafted graphs. 


\noindent\textbf{Query-based attacks.} Based on the type of query response accessible to the adversary, existing attacks operate under either soft-label settings, where the victim returns full probability 
distributions over classes, or hard-label settings, where only the top-1 predicted labels are available. Shen et al. \cite{IGNN_MSA_2022_SP} propose the first systematic study of model stealing attacks against inductive GNNs, where the adversary queries the victim with shadow graphs from the same distribution as the training graph and exploits posterior probabilities, node embeddings, and t-SNE projections to train the surrogate. Their follow-up work~\cite{IGNN_MSA_2024} further introduces graph contrastive learning and spectral data augmentation to enhance attack performance. However, both methods require access to a substantial shadow dataset, i.e., 30\% of the victim training graph, and assume the availability of rich intermediate outputs that are rarely accessible in real-world MLaaS deployments. CEGA \cite{wang2025cega} proposes an iterative node selection framework that adaptively queries nodes over multiple cycles using historical feedback. While CEGA reduces query cost, its multi-cycle active querying paradigm requires global graph topology to compute PageRank-based centrality \cite{Page1999ThePC}, relies on K-means over global node embeddings to measure diversity, and depends on soft-label outputs to estimate prediction uncertainty. Furthermore, its sequential and high-frequency query patterns can easily trigger system-level anomaly detectors \cite{prada}.

Under the strict hard-label setting, DeFazio et al. \cite{GNN_MSA_2019} pioneer GNN model stealing attacks by introducing graph perturbation techniques on 2-hop subgraphs of each target node to synthesize queries. They empirically find that training the surrogate model with top-1 hard labels yields even higher fidelity than utilizing soft-label outputs. 
Wu et al. \cite{GNN_MSA_2022_Asia_CCS} further leverage diverse background knowledge, such as partial features or topologies, or full shadow graphs to synthesize missing information for inaccessible nodes. 
Guan et al. \cite{RealisticGNN_MSA_2024_KBS} additionally introduce an edge prediction module to mitigate noise propagation from incorrect predicted labels. However, their method still assumes the presence of relatively connected graph topologies and requires 25\% of the training graph. 


\noindent\textbf{Data-free attacks.} Zhuang et al. \cite{DF_GNN_MSA_2024_USENIX} train a GAN-style graph generator to synthesize query graphs, which are subsequently used to query the victim and train the surrogate. Despite being data-free, this GAN-style attack framework introduces substantial computational overhead, training instability and high detection risks, while still requiring a massive query budget through iterative generator-victim interactions.

\noindent\textbf{Limitations of prior work.}
Despite significant progress, existing GNN stealing attacks share several critical bottlenecks that hamper their real-world applicability:
(1) \textbf{Data dependency:} most methods assume access to large amounts of or even full victim training graphs even under soft-label settings;
(2) \textbf{Structural over-reliance:} acquiring complete multi-hop subgraphs for each target node is often impractical, as it causes unintended topological and feature leakage that exceeds realistic API access permissions; 
(3) \textbf{High query overhead and detection risks:} query budgets remain prohibitively high for strict black-box deployments, where excessive queries not only incur significant monetary costs but also risk triggering query-based anomaly detection deployed by MLaaS providers;
(4) \textbf{Backbone Reliance:} existing GNN stealing attacks commonly assume prior knowledge of the victim's backbone, which rarely holds in realistic black-box deployments as such knowledge is proprietary and closely guarded by service providers;
(5) \textbf{Evaluation inconsistency:} with few exceptions in inductive settings, open-source implementations of the remaining methods indicate that they implicitly query the victim with the full graph while only returning the query responses of target nodes, creating a fundamental discrepancy between claimed threat models and experimental evaluations.
In contrast, our work operates under a strictly more constrained yet realistic threat model: \textit{black-box hard-label access only}, \textit{queries restricted to the induced subgraphs of target nodes} without full-graph access or complete subgraph information, and with a \textit{limited budget of only 5\% of the victim training graph}.

\section{Threat Model and Empirical Studies}
\label{sec:prelim_exp}

In this section, we first formalize the threat model and problem statement, establishing a more restricted yet realistic attack setting than prior GNN stealing work. We then conduct three empirical studies to reveal the unique challenges arising from this restricted attack setting. Together, these challenges motivate the holistic design of our framework in Section \ref{sec:attack_framework_design}.

\subsection{Threat Model}

\subsubsection{\textbf{Attack Goals}}
In line with the standard taxonomy \cite{IGNN_MSA_2022_SP, ori_taxonamy, IGNN_MSA_2024}, we consider two adversarial objectives: theft and reconnaissance. A \textbf{theft} adversary aims to construct a surrogate model $\Phi_S$ that achieves comparable task accuracy to the victim model $\Phi_V$, thereby compromising intellectual property and confidentiality. A \textbf{reconnaissance} adversary instead seeks to replicate the victim's model behaviors. Ideally, $\Phi_S$ should agree with $\Phi_V$ on any given input, including cases where both models make incorrect predictions. This alignment is quantified by fidelity, where a high-fidelity surrogate serves as a springboard for downstream exploitation, such as crafting transferable adversarial examples. 

\subsubsection{\textbf{Attacker's Knowledge and Capabilities}}
We consider a strict yet realistic black-box and hard-label attack setting. The adversary can interact with the victim solely through a query interface, such as a remotely accessible API, and receive hard-label responses. No inner parameters, backbone prior knowledge, architectural configurations, training procedures or intermediate representations are available.

Unlike prior work that explicitly or implicitly relies on querying full graphs, we strictly restrict query access to the target nodes and their induced subgraph. To align with standard settings that mirror real-world MLaaS deployments \cite{GNN_MSA_2019, GNN_MSA_2022_Asia_CCS, IGNN_MSA_2022_SP, IGNN_MSA_2024}, the adversary can submit $k$-hop local subgraphs of target nodes as query inputs. However, distinct from prior formulations where adversaries can acquire complete multi-hop subgraphs derived from the full victim training graph, our adversary can only construct query graphs within the induced subgraph. On sparse or small-scale graphs, such queries frequently collapse into sets of isolated nodes. This constraint not only prevents the adversary from exploiting global graph structure or cross-boundary neighborhood information that would be inaccessible in practice, but also reduces the risk of triggering rate limits or anomaly detectors \cite{prada}.

The attack budget is strictly bounded by $5\%$ of the victim training graph. We further prohibit data synthesis to expand queries, as synthetic samples lie outside the natural data manifold and are more vulnerable to distribution-aware defenses \cite{prada}. In contrast, querying only real training nodes renders queries indistinguishable from legitimate usage.


\subsubsection{\textbf{Attack Scenarios}}
We outline three representative real-world attack scenarios as follows:

\textbf{(1) Commercial MLaaS Model Theft:} A competitor company/institution queries a commercial GNN API to replicate its underlying functionality at minimal cost, bypassing proprietary data collection and intensive model training.

\textbf{(2) Low-footprint Adversarial Transfer:} An adversary first extracts a surrogate under a strict query budget to evade rate-limiting and anomaly defenses, then crafts transferable adversarial examples to attack the victim \cite{IGNN_MSA_2022_SP}.

\textbf{(3) Intellectual Property Audit:} An auditor extracts a surrogate model under harsh constraints to assess model vulnerability or verify unauthorized IP infringement, providing critical insights for downstream defense mechanisms.

\subsection{Problem Statement}
\label{ps}

Based on the threat model described above, we now formalize our problem as follows.

Given $\mathcal{G}\!=\!(\mathcal{V}, \mathcal{E}, X)$, a victim GNN 
$\Phi_V: \mathcal{G} \rightarrow \mathbb{R}^{|\mathcal{V}| \times C}$ is trained on $\mathcal{G}$ for the node classification task over $C$ classes and deployed as a black-box service. The adversary has access to the target node set $\mathcal{V}_q \subseteq \mathcal{V}$ subject to a budget constraint:
\begin{equation}
    |\mathcal{V}_q| \leq \beta \cdot |\mathcal{V}|
\end{equation}
where $\beta$ is the graph access ratio, we set $\beta=5\%$.

Notably, the induced subgraph $\mathcal{G}_q$ only covers inner connections among these target nodes $\mathcal{V}_q$, and it can be largely unconnected with many isolated nodes:
\begin{equation}
    \mathcal{G}_q = \mathcal{G}[\mathcal{V}_q] = 
    \left(\mathcal{V}_q,\; 
    \mathcal{E} \cap (\mathcal{V}_q \times \mathcal{V}_q),\;
    X_{\mathcal{V}_q}\right)
\end{equation}

For each queried node $v \in \mathcal{V}_q$, the adversary submits its k-hop local subgraph within the induced subgraph, i.e.,  $\mathcal{G}_v^k\subseteq \mathcal{G}_q$ (with $k\!=\!2$ throughout) as the query input, and the victim returns the hard-label prediction $\hat{y}_v^V$ for the target node. 
\begin{equation}
    \mathcal{G}_v^k = \mathcal{G}_q\!\left[
    \mathcal{N}_{\mathcal{G}_q}^k(v) \cup \{v\}
    \right]
\end{equation}
\begin{equation}
    \hat{y}_v^V = \arg\max_{c \in \mathcal{C}}\; 
    \left[\Phi_V\!\left(\mathcal{G}_v^k\right)\right]_{v,c}
\end{equation}

Given only the query-response pairs $\mathcal{D}\!=\!\{(\mathcal{G}_v^k,\, \hat{y}_v^V)
\}_{v \in \mathcal{V}_q}$ collected under the above constraints, the adversary seeks to train a surrogate model $\Phi_S$ that simultaneously achieves theft and reconnaissance goals against ground-truth and predicted labels on unseen nodes $\mathcal{V}_{new}$:
\begin{equation}
    \text{Acc}(\Phi_S) = 
    \frac{1}{|\mathcal{V}_{new}|}
    \sum_{v \in \mathcal{V}_{new}} 
    \mathbf{1}[\hat{y}_v^S = y_v]
\end{equation}
\begin{equation}
    \text{Fid}(\Phi_S) = 
    \frac{1}{|\mathcal{V}_{new}|}
    \sum_{v \in \mathcal{V}_{new}} 
    \mathbf{1}[\hat{y}_v^S = \hat{y}_v^V]
\end{equation}
where $y_v$ is the ground-truth label of node $v$. The inherent bottleneck is that the surrogate $\Phi_S$ must generalize to unseen nodes, despite being trained only on hard labels derived from the potentially sparse and small-scale induced subgraph $\mathcal{G}_q$. 

\subsection{Empirical Studies}
Before detailing our attack design, we conduct three empirical studies to investigate the unique challenges imposed by our strict threat model across four representative real-world graph datasets: Cora, PubMed \cite{cora_pubmed}, Amazon-Computer (Computer, hereinafter) and Physics \cite{shchur2019pitfallsgraphneuralnetwork}. Cora and PubMed are computer science and biomedical citation networks, respectively, where each node denotes a document with a bag-of-words representation, and each edge represents a citation link. Computer and Physics are built upon Amazon co-purchase and academic collaboration networks, where each node signifies a product or an author, each node feature vector encodes a product review or paper keywords, and each edge indicates co-purchases or co-authorship. Dataset statistics are summarized in Table \ref{graph_data}. 

\begin{table}[H]
\caption{Statistics of the Graph Datasets}
\label{graph_data}
\resizebox{\linewidth}{!}{
\begin{tabular}{c|c|c|c|c|c|c}
\toprule
\textbf{Dataset}  & \textbf{\# of Nodes}   & \textbf{\# of Edges}    & \textbf{\# of Features}  & \textbf{\# of Classes}  & \textbf{Avg. Degree} & \textbf{Node Homophily (\%)} \\
\hline
\textbf{Cora}     & 2,708  & 10,556  & 1,433 & 7 & 3.90    & 82.52           \\
\textbf{PubMed}   & 19,717 & 88,651  & 500  & 3 & 4.50    & 79.24           \\
\textbf{Computer} & 13,752 & 491,722 & 767  & 10 & 35.76   & 78.53           \\
\textbf{Physics}  & 34,493 & 495,924 & 8,415 & 5 & 14.38   & 91.53  \\ \bottomrule     
\end{tabular}}
\end{table}

\subsubsection{Label Quality Degradation} 
Under the threat model defined above, the adversary's query access is doubly constrained: the query budget is strictly limited, and queries are fully restricted to the induced subgraph of the target nodes rather than the complete subgraph in the victim training graph. As a result, the compound effect of these two constraints gives rise to highly sparse subgraphs and even isolated nodes, particularly on sparse or small-scale graphs. This raises a natural question: can GNNs reliably perform inference on such degenerate graph structures? To investigate this, we evaluate GNN prediction quality on 5\% randomly sampled target nodes under four topology conditions: isolated nodes (features only), 1-hop and 2-hop induced subgraphs, and the full graph, across all benchmark graphs and diverse GNN backbones.


\begin{figure}[htbp]
    \centering
    \includegraphics[width=1\linewidth]{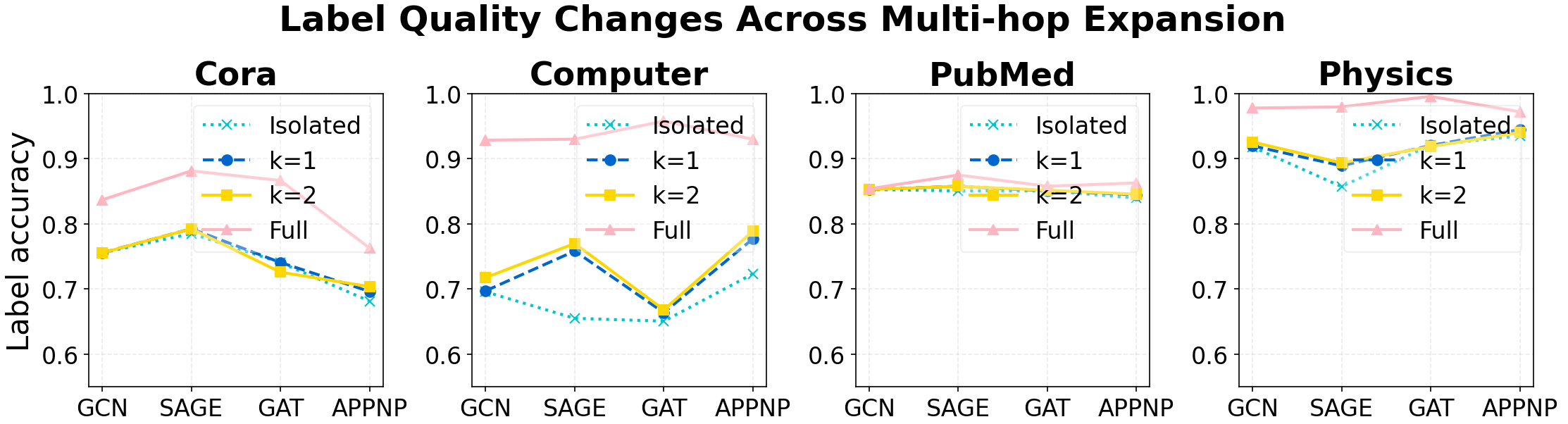}
    \caption{Label Quality Changes across Multi-hop Expansion}
    \label{label_quality}
\end{figure}


\noindent\textbf{Empirical Results.} Figure \ref{label_quality} shows that label accuracy generally degrades as the available topology decreases. On relatively sparse graphs such as Cora and PubMed, incrementally expanding topological information does not consistently improve label quality. In fact, 2-hop induced subgraphs occasionally yield lower label accuracy than isolated nodes, which indicates that most 1-hop and 2-hop neighbors within the induced subgraphs belong to different classes, thereby introducing noise that distorts victim predictions. 

\noindent\textbf{Takeaways.} The above empirical results reveal two fundamental challenges imposed by the strict threat model to design GNN stealing attacks:

(1) \textbf{C1: Poor Surrogate:} the highly sparse induced subgraph with a large number of isolated nodes severely degrades standard GNN message passing mechanism, with few or no neighbors to aggregate, GNNs fail to learn informative node representations that should encapsulate both topological and feature context. This representation collapse fundamentally misaligns with the latent space of the victim model trained on the full graph, leaving the surrogate incapable of faithfully approximating the victim's behaviors. 

(2) \textbf{C2: Insufficient Supervision:} hard-label queries inherently discard inter-class similarity information encoded in soft probabilities, leaving the surrogate with insufficient training signals for generalization. To make matters worse, the victim predictions on isolated nodes and incomplete subgraphs are substantially less accurate than under full-graph inference, further compromising supervision reliability. Together, these factors render the query-response pairs both informationally encapsulated and noisy, posing a fundamental obstacle to surrogate model training. 

\subsubsection{Class Imbalance and Absence}
Beyond degrading prediction quality, the sparse structures and isolated nodes identified above introduce a further complication: the resulting label distribution is highly imbalanced, and certain classes may be absent entirely from the queried nodes. We refer to this as the class imbalance and absence problem, and investigate its impact on GNN stealing performance below.

To quantify class imbalance, we define the imbalance ratio as the proportion of samples in the most frequent relative to the least frequent class over ground-truth or predicted labels. 

\begin{figure}[htbp]
    \centering
    \includegraphics[width=1\linewidth]{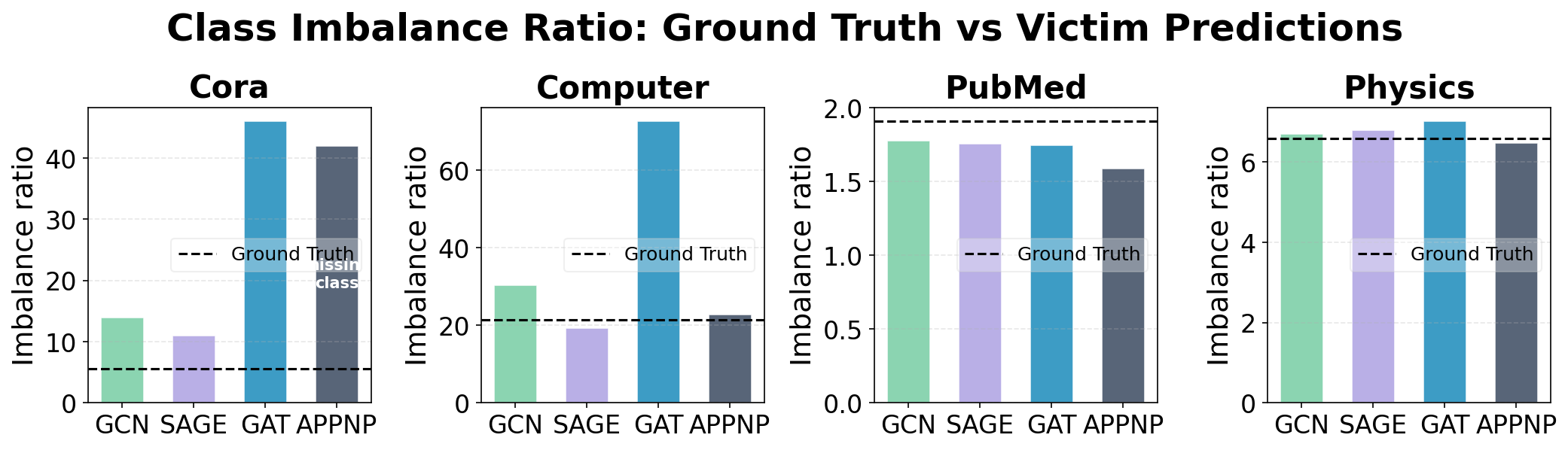}
    \caption{Class Imbalance Ratio}
    \label{class_ir}
\end{figure}

\begin{figure}[htbp]
    \centering
    \includegraphics[width=1\linewidth]{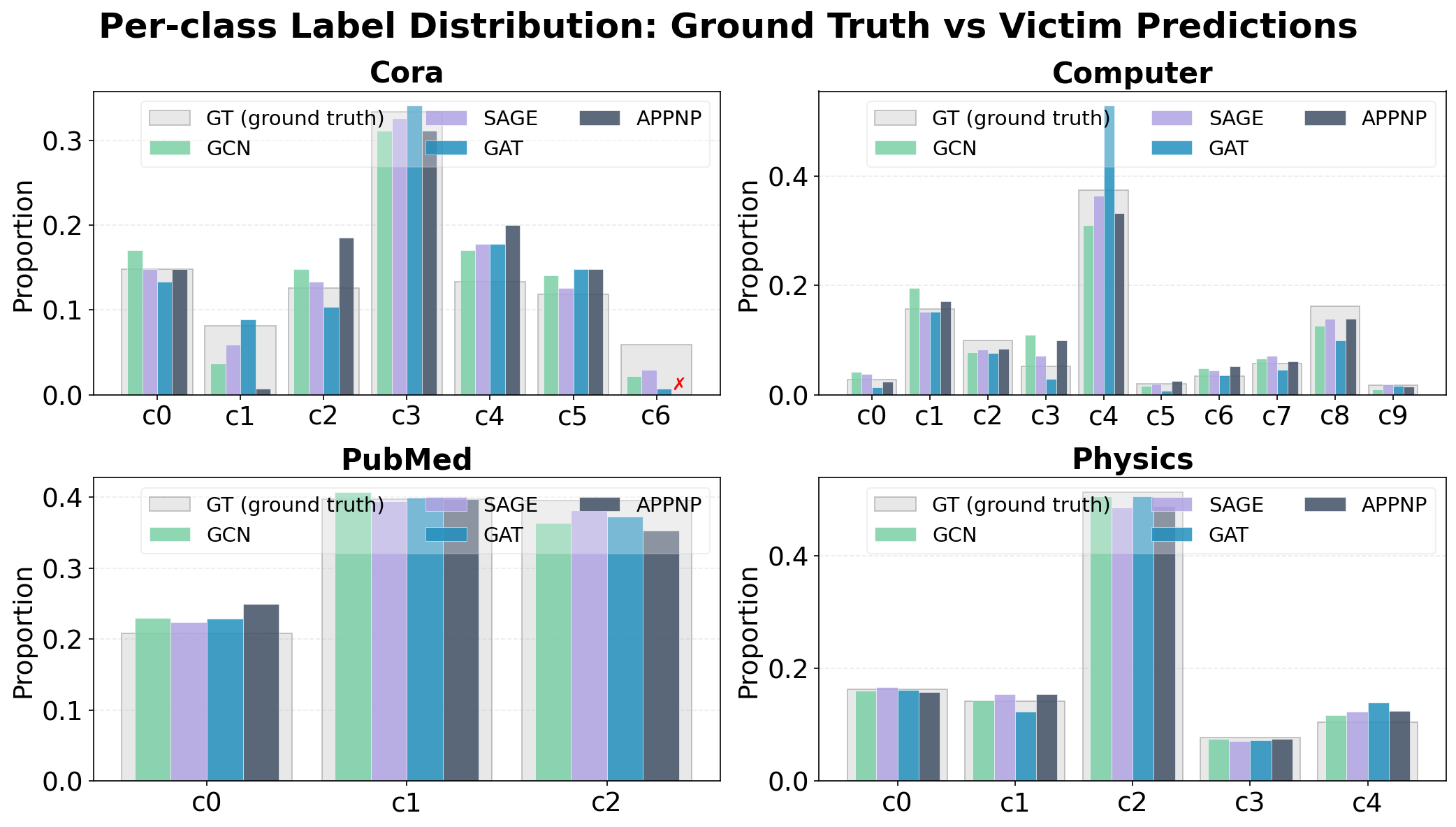}
    \caption{Per-class Label Distribution}
    \label{dist_per_class}
\end{figure}

\noindent\textbf{Empirical Results.} As illustrated in Figure \ref{class_ir} and \ref{dist_per_class}, victim predictions on the induced subgraphs systematically amplify class imbalance relative to the ground-truth distribution, with imbalance ratios up to 45$\times$ on Cora and 70$\times$ on Computer when using GAT as the victim backbone, far exceeding the ground-truth baselines. In contrast, PubMed displays a relatively balanced prediction distribution, where victim imbalance ratios remain slightly below the ground-truth baseline across all backbones. While variations among victim backbones retain minor on Physics, they become substantial on other graphs with larger class space. More critically, when using APPNP as the victim backbone on Cora, class 6 is missing and class 1 only contains a single target node. Such extreme class imbalance and missing classes severely bias the surrogate model toward majority classes, leaving sparse or absent training signals for minority classes.

\noindent\textbf{Takeaways.} These empirical findings highlight another fundamental challenge--\textbf{C3: Class Imbalance and Absence}. Querying hard labels on sparse induced subgraphs inherently exacerbates class imbalance: structurally dominant classes capture a disproportionate share of target predictions, while minority classes remain severely under-represented or entirely absent from the query set. This systematic bias consequently skews the surrogate model's classifier toward majority classes, further degrading stealing performance beyond what structural sparsity alone would cause. 

\subsubsection{Backbone Disagreement}

\begin{figure}[htbp]
    \centering
    \includegraphics[width=1\linewidth]{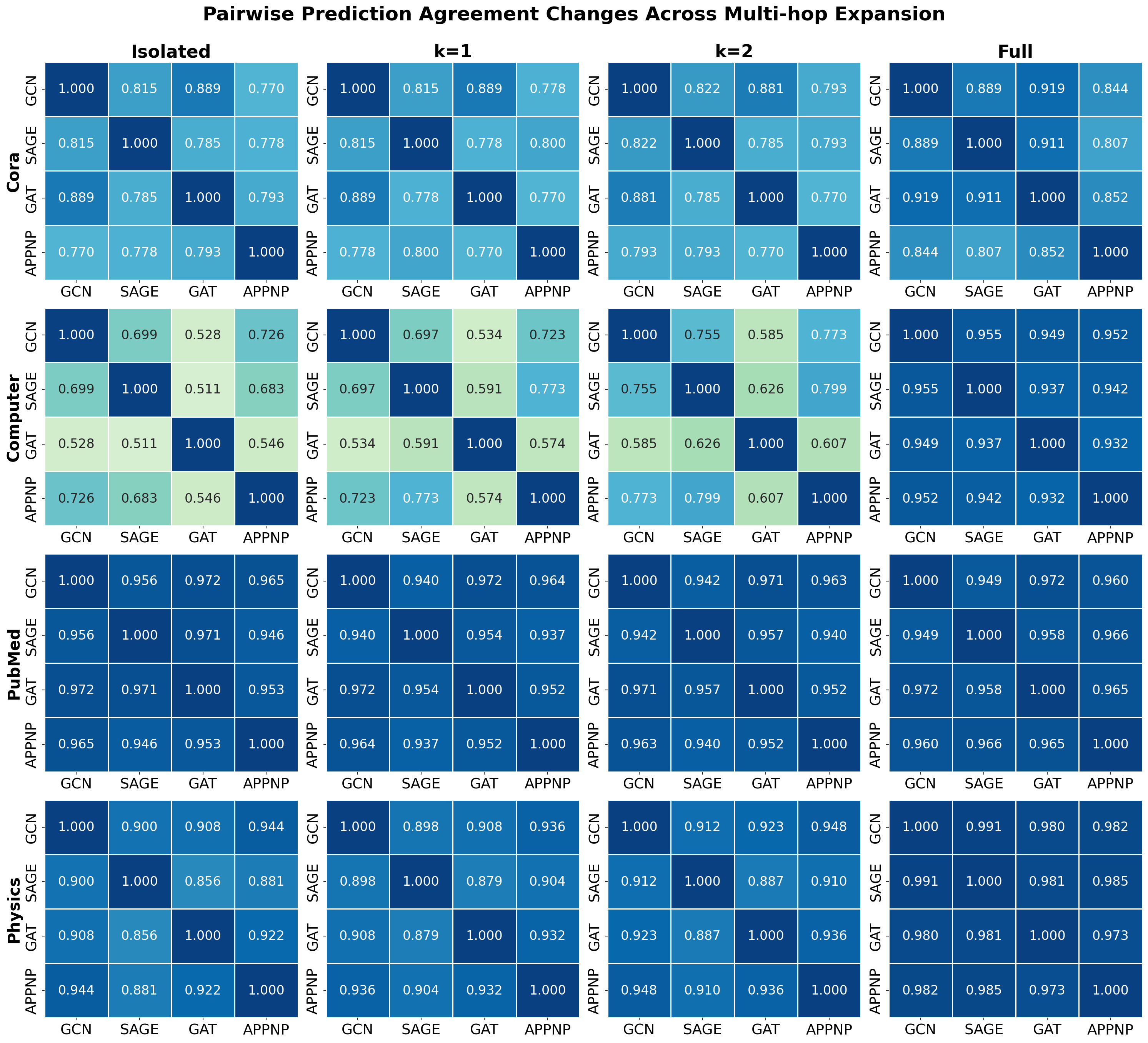}
    \caption{\centering Pairwise Prediction Agreement Changes across Multi-hop Expansion}
    \label{pair_diagree}
\end{figure}

Existing GNN model stealing attacks typically assume that the adversary masters the knowledge of victim backbone \cite{GNN_MSA_2019, GNN_MSA_2022_Asia_CCS, IGNN_MSA_2022_SP, wang2025cega}. However, this information is rarely available in practice due to commercial intellectual property protection and the black-box nature of MLaaS APIs. To understand the impact of victim backbone misalignment, we evaluate the pairwise prediction agreement across different victim backbones under our strict setting and present the results in Figure \ref{pair_diagree}. 

\noindent\textbf{Empirical Results.} We observe that the victim prediction agreement is substantially lower on sparse induced subgraphs than on the full graph, with off-diagonal entries as low as 0.51 on Computer for isolated nodes. This indicates that different backbones exhibit divergent behaviors as structural information decreases, making cross-backbone model stealing inherently more challenging under our restricted threat model. The progressive increase in agreement from isolated to full graphs confirms that topology serves as the primary driver of backbone divergence.

\noindent\textbf{Takeaways.} These observations underscore a key challenge--\textbf{C4: Backbone Mismatch.} Backbone mismatch further compounds the aforementioned challenges, as different backbones exhibit varying degrees of prediction degradation under the same induced subgraph queries. However, our threat model assumes a practical black-box adversary with no prior knowledge of the victim's backbone, necessitating an attack framework that remains strictly backbone-agnostic.

\section{Attack Framework Design}
\label{sec:attack_framework_design}

In this section, we propose Dagger, a novel decoupling-based attack framework tailored to the four challenges identified in Section \ref{sec:prelim_exp}. Specifically, \textbf{C1 (Poor Surrogate)} demands that an effective surrogate must remain topology-adaptive, preserving rich semantic context for isolated nodes while extracting structural information across subgraphs of varying scale and density; \textbf{C2 (Insufficient Supervision)} requires maximizing the utility of collected query-response pairs while filtering or smoothing unreliable predictions to prevent error propagation; \textbf{C3 (Class Imbalance and Absence)} necessitates incorporating class-aware recalibration to mitigate systematic biases and ensure balanced classifier optimization across minority classes; and \textbf{C4 (Backbone Mismatch)} calls for the attack framework to remain strictly backbone-agnostic.

\begin{figure}
    \centering
    \includegraphics[width=1\linewidth]{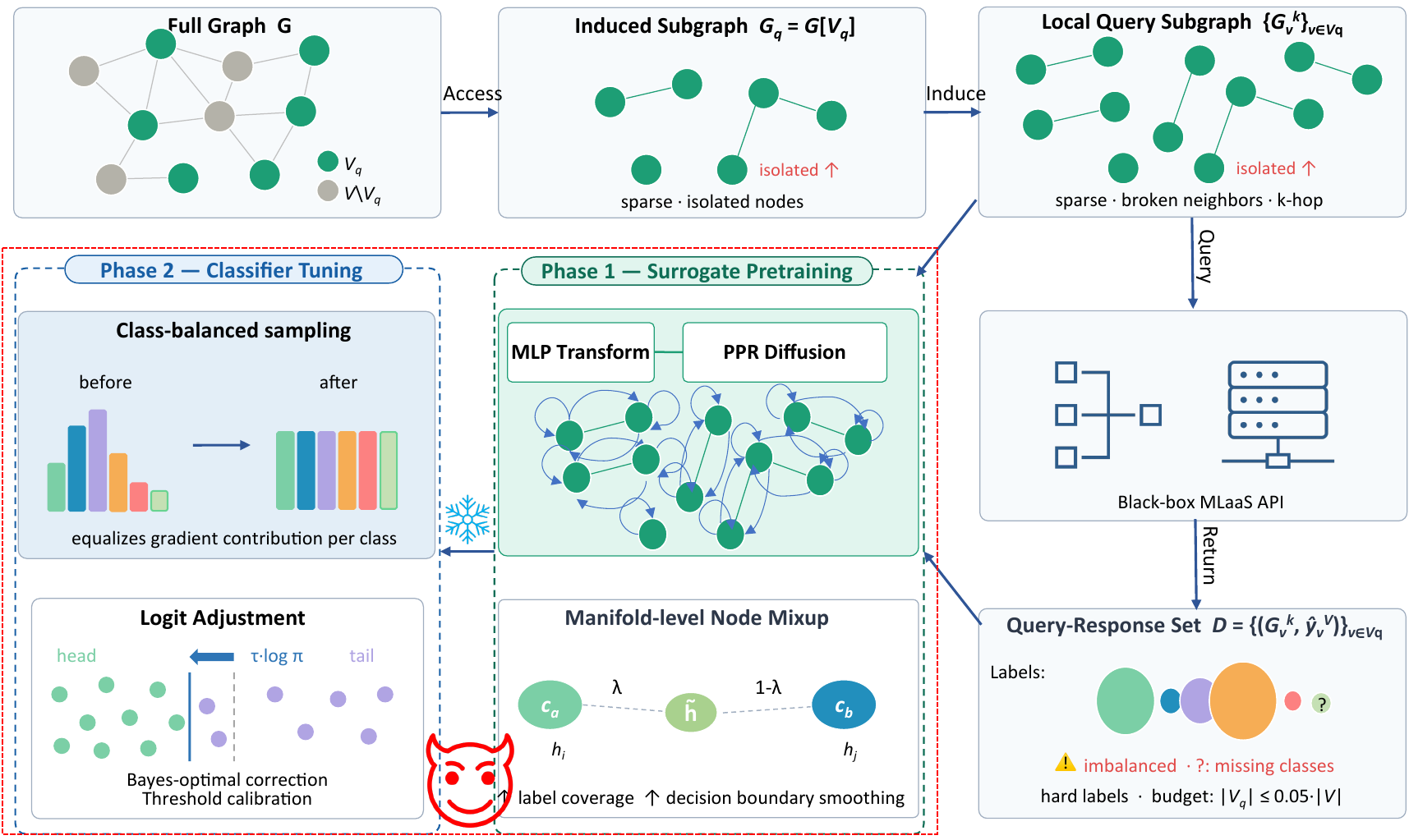}
    \caption{The Overview of Dagger}
    \label{dagger}
\end{figure}

As illustrated in Figure \ref{dagger}, Dagger proceeds in two phases. In \textbf{Phase 1}, to tackle C1 and C2, we train a surrogate encoder via decoupling-based feature propagation and manifold-level mixup. This design mitigates the structural collapse of sparse induced subgraphs and enriches context-deficient node representations without over-relying on message passing. 
In \textbf{Phase 2}, to overcome C3, we freeze the pre-trained encoder and tune only the classifier head using class-balanced sampling paired with logit adjustment, rectifying predicted-label class imbalance and ensuring backbone-agnostic generalization. 
Notably, by decoupling the encoder from the classifier and avoiding any assumption about the victim's backbone knowledge throughout both phases, Dagger inherently satisfies the backbone-agnostic requirement imposed by C4.

\subsection{Phase 1: Decoupling-based Surrogate Pretraining}

\subsubsection{Topology-adaptive Surrogate (C1)}

Due to the double constraints of query access under our threat model, the query subgraphs $\{\mathcal{G}_v^k\}_{v \in \mathcal{V}_q}$ suffer from three compounding structural deficiencies. First, they can be highly sparse and frequently unconnected, particularly on sparse or small-scale graphs, causing standard message-passing GNNs trained on such degenerative structures to degrade towards Multi-Layer Perceptrons (MLPs). Second, nodes with originally high degrees may become isolated once their neighbors are excluded from the target node set, causing GNNs to lose the ability to extract information from their multi-hop neighbors. Third, even among connected nodes within $\{\mathcal{G}_v^k\}_{v \in \mathcal{V}_q}$, their scope of information propagation is confined to a small local range, which makes it difficult to integrate broader structural context. 

These structural deficiencies motivate a topology-adaptive surrogate that simultaneously handles both isolated nodes and highly-sparse local structures without over-relying on message passing. To this end, we adopt the decoupling design of APPNP \cite{appnp}: predict and then propagate, separating feature transformation from graph propagation into two explicit steps. For each target node $v$, we first utilize an MLP to extract the initial representation $H_v^{(0)}$ from raw features in local subgraphs $\mathcal{G}_v^k$, independent of its local topology. This prediction step obtains a meaningful representation for isolated nodes rather than collapsing to uninformative embeddings while laying the foundation for subsequent structural information propagation among connected ones.
\begin{equation}
    H_v^{(0)} = \text{MLP}(X_{\mathcal{G}_v^k})
\end{equation}

We then leverage Personalized PageRank (PPR) \cite{ppr,appnp} to propagate $H^{(0)}_v$ across $K$ steps, retaining an $\alpha$ fraction of the original representation at each step to adaptively incorporate local structural context while prevent over-smoothing: 
\begin{equation}
    H_v^{(k)} = (1-\alpha)\hat{A}_v H_v^{(k-1)} 
    + \alpha H_v^{(0)}, \quad k = 1, \ldots, K
\end{equation}
where $\hat{A}_v$ is the adjacency matrix of $\mathcal{G}_v^k$ with 
self-loops. Mechanistically, $H_v^{(k)}$ acts as an information origin while $\alpha$ serves as a restart probability to avoid diluting personalized representations over multiple hops, thereby effectively mitigating over-smoothing and enabling deep information propagation. 

Standard message-passing GNNs implicitly rely on the node homophily assumption that connected nodes tend to share similar features and structural patterns \cite{node_homo}. To accommodate isolated nodes during PPR diffusion, we explicitly add self-loops to these nodes. Since a node exhibits maximum similarity with itself, the addition of self-loops transforms the diffusion mechanism into a self-reinforcing process that guarantees stable feature propagation and prevents representation collapse.

\subsubsection{Knowledge Distillation Loss}
Given the query-response pairs $\mathcal{D}\!=\!\{(\mathcal{G}_v^k,\, \hat{y}_v^V)
\}_{v \in \mathcal{V}_q}$, we train the surrogate by minimizing the cross-entropy loss between its predictions on the target node $v$ and the victim's hard-label predictions:
\begin{equation}
    \mathcal{L}_{\text{distill}} = 
    -\frac{1}{|\mathcal{V}_q|}
    \sum_{v \in \mathcal{V}_q}
    \log \frac{
        \exp\!\left(\left[\Phi_S(\mathcal{G}_v^k)
        \right]_{v,\,\hat{y}_v^V}\right)
    }{
        \sum_{c \in \mathcal{C}} 
        \exp\!\left(\left[\Phi_S(\mathcal{G}_v^k)
        \right]_{v,\,c}\right)
    }
\end{equation}
where $\left[\Phi_S(\mathcal{G}_v^k)\right]_{v}$ denotes the logit vector of the center node $v$ in the output of the surrogate $\Phi_S$.

\subsubsection{Manifold Mixup Regularization (C2)}
A further challenge arises from incomplete class coverage: queries restricted to local subgraphs $\{\mathcal{G}_v^k\}_{v \in \mathcal{V}_q}$ may fail to elicit predictions for all $|\mathcal{C}|$ classes from the victim model, particularly on sparse or small-scale graphs with multiple classes. Beyond this, hard-label supervision discards inter-class relationships encoded in the victim's soft-label predictions. Training on such sparse supervision forces the surrogate to overfit to the limited query space and learn spurious decision boundaries, severely hampering its generalization to unseen nodes.

To bridge these gaps, we apply manifold-level node mixup to synthesize continuous interpolated representations across class boundaries. We finally smooth the decision boundary in the manifold rather than the input space. Interpolating raw node features is ill-defined for graph-structured data: node features are typically sparse and discrete (e.g., bag-of-words), rendering their linear combinations semantically meaningless, while input-space interpolation disregards the underlying non-Euclidean graph topology. In contrast, manifold interpolation directly operates on a continuous latent space that inherits the structural context encoded by PPR propagation and encourages the surrogate to learn more transferable representations. 

We then formalize the manifold-level node mixup. For each query node $v_i$ with predicted-label $\hat{y}_{v_i}^V\!=\!c_a$, we sample another query node $v_j$ with a different predicted-label $\hat{y}_{v_j}^V\!=\!c_b$, where $(v_i, \!c_a)$ and $(v_j, \!c_b)$ both belong to the query-response pairs $\mathcal{D}\!=\!\{(\mathcal{G}_v^k,\, \hat{y}_v^V)
\}_{v \in \mathcal{V}_q}$, $i\neq j, c_a\neq c_b$, and interpolate their focal-node hidden representations $H_{v_i}$ and $H_{v_j}$:
\begin{equation}
    \tilde{H}_{v_{ij}} = \lambda \cdot H_{v_i} + 
    (1-\lambda) \cdot H_{v_j}, 
    \quad \lambda \sim \text{Uniform}(0,\, 1)
\end{equation}
Unlike the feature-level node mixup \cite{node_mixup} which employs Beta distributions to concentrate mixing ratios near the original samples, we adopt $\lambda \sim \text{Uniform}(0, 1)$ to uniformly explore the full interpolation spectrum between class boundaries. This setup particularly aligns with our attack setting: since node representations reside on a continuous semantic manifold rather than a discrete input space, dense and uniform sampling across $(0, 1)$ synthesizes a richer variety of intermediate representations, providing more comprehensive coverage of the decision boundary between $c_a$ and $c_b$.

The mixed representation $\tilde{H}_{v_{ij}}$ is trained against a paired soft label $\tilde{y}_{ij} = \lambda \cdot \mathbf{e}_{c_a} + 
(1-\lambda) \cdot \mathbf{e}_{c_b}$ via KL divergence:
\begin{equation}
    \mathcal{L}_{\text{mix}} = 
    \frac{1}{|\mathcal{V}_q|}
    \sum_{v_i \in \mathcal{V}_q}
    D_{\text{KL}}\!\left(
        \tilde{y}_{ij} \;\big\|\; 
        \text{softmax}\!\left(
        f_{S}(\tilde{H}_{v_{ij}})
        \right)
    \right)
\end{equation}
where $\mathbf{e}$ is a one-hot vector and $f_{S}$ denotes the linear classifier head of $\Phi_S$. The soft labels $\tilde{y}_{ij}$ further complement inter-class relationships that are discarded by hard-label supervision.

\subsubsection{Phase Objective}
The final objective of Phase 1 is:
\begin{equation}
    \mathcal{L}_{\text{Phase 1}} = 
    \mathcal{L}_{\text{distill}} + 
    w_{\text{mix}} \cdot \mathcal{L}_{\text{mix}}
\end{equation}
where the coefficient $w_{\text{mix}}$ controls the contribution of manifold-level node mixup.

\subsection{Phase 2: Imbalance-aware Classifier Tuning}

After pretraining in Phase 1, the surrogate has learned a representation space that captures the victim's functional behaviors. While manifold-level node mixup mitigates the class imbalance by implicitly augmenting minority class representations, it operates at the representation level and does not directly correct the skewed decision boundaries induced by majority-class dominance in the training distribution. As a result, the classifier head remains systematically biased towards majority-class nodes, leaving minority classes under-represented and consequently eroding the overall fidelity of the surrogate. To rectify this bias, Phase 2 freezes the surrogate encoder $\text{Enc}_{S}(\cdot)$ to preserve the learned representations and only fine-tunes the linear classifier head $f_{S}$ with two complementary corrections: class-balanced sampling and logit adjustment \cite{longtail_la}.

\subsubsection{Class-balanced Sampling (C3)} To counteract the majority class dominance in the query-response pairs  $\mathcal{D}$, we oversample minority classes with replacement such that each class provides exactly the same number of nodes per epoch as the majority class. The sampling probability is defined as:
\begin{equation}
    p(v) = \frac{1}{|C| \cdot n_c}, 
    \quad v \in \mathcal{V}_q^{(c)}
\end{equation}
where $n_c\!=\!|\mathcal{V}_q^{(c)}|$ is the number of query nodes for class $c$.

This sampling strategy balances per-class gradient contributions, preventing the classifier from collapsing to majority-class predictions and ensuring that minority classes receive sufficient gradient signals throughout fine-tuning.

\subsubsection{Logit Adjustment (C3)}
Class-balanced sampling equalizes per-class gradient contributions and stabilizes optimization. However, oversampling minority classes inherently alters the class prior from the skewed query distribution to a uniform one, inevitably introducing a distribution shift. To mitigate this discrepancy without resorting to complex optimization, we adopt a simple-yet-effective technique called logit adjustment \cite{longtail_la} to calibrate the classifier logits proportional to $\log \pi_c$:
\begin{equation}
    \tilde{f}_S(H)_c = f_S(H)_c - \tau \cdot \log \pi_c
\end{equation}
where $\pi_c\!=\!n_c / |\mathcal{V}_q|$, $c\in C$ denotes the empirical class prior estimated from the victim-predicted label distribution, while $\tau\!>\!0$ is a temperature parameter controlling the strength of the adjustment. This technique is theoretically grounded: Menon et al. \cite{longtail_la} show that under the adjusted loss, the optimal classifier recovers the Bayes-optimal decision rule for the balanced distribution, effectively redistributing decision boundaries from majority to minority classes without requiring additional queries to the victim.

\subsubsection{Phase Objective}
The Phase 2 training objective is:
\begin{equation}
    \mathcal{L}_{\text{Phase 2}} = 
    -\frac{1}{|\mathcal{V}_q|}
    \sum_{v \in \mathcal{V}_q}
    \log \frac{
        \exp\!\left(\tilde{f}_S(H_v^S)_{\hat{y}_v^V}\right)
    }{
        \sum_{c=1}^{C} 
        \exp\!\left(\tilde{f}_S(H_v^S)_c\right)
    }
\end{equation}
where $H_v^S\!=\!\text{Enc}_S(v)$ denotes the frozen focal-node embedding from Phase 1. These two corrections are complementary by design: class-balanced sampling stabilizes training by equalizing per-class gradient contributions, while logit adjustment restores proper calibration during inference by explicitly compensating for the prior shift induced by rebalancing.

\section{Experiments}
\label{sec:exp}
In this section, we conduct empirical experiments to investigate the following key research questions:

\noindent\textbf{RQ1}: How effective and efficient is Dagger to launch model stealing attacks against GNNs?

\noindent\textbf{RQ2}: Can Dagger flexibly adapt to different GNN backbones?

\noindent\textbf{RQ3}: How Dagger performs under diverse mismatched architecture configurations?

\noindent\textbf{RQ4}: What factors affect the attack performance of Dagger?

\noindent\textbf{RQ5}: How does each component contribute to Dagger?

\noindent\textbf{RQ6}: Can Dagger bypass diverse defense mechanisms?


\subsection{Setups}

\noindent\textit{\textbf{GNN Backbones}}
We employ four GNN backbones: GCN \cite{gcn}, GAT \cite{gat}, APPNP \cite{appnp}, and GraphSAGE \cite{sage}, and report their benign performance across aforementioned four graphs in Table \ref{benign_gnn} in Appendix \ref{sec_benign}.

\noindent\textit{\textbf{Baselines.}}
To quantify the contribution of real graph structural information to attack performance, we introduce a synthetic graph baseline (Rand.) that requires no access to real graph data. We generate a Stochastic Block Model (SBM) graph with $|\mathcal{C}|$ equal-sized blocks, intra-block edge probability $p_{\text{intra}}\!=\!0.3$, and inter-block edge probability $p_{\text{inter}}\!=\!0.05$, with the total number of nodes matching the $5\%$ query budget of each dataset. Node features are constructed by randomly sampling rows from the real graph's feature matrix with replacement and adding small Gaussian perturbations $\mathcal{N}(0, 0.01^2)$, preserving the marginal feature distribution of the target dataset while decoupling the feature-topology correlations present in the real graph. We further compare Dagger with the state-of-the-art (SOTA) GNN stealing attacks, under the same access to $5\%$ of the victim training graph, including 1) MEA-Attack0 \cite{GNN_MSA_2022_Asia_CCS}: to align with its setting, the adversary can obtain 2-hop neighbors of each target node. She/he then synthesizes the neighboring attributes via homophily-based neighbor expansion. 2) AdvMEA \cite{GNN_MSA_2019}: the adversary constructs fully-connected synthetic subgraphs by sampling node features from per-class empirical distributions from the whole graph. 3) CEGA \cite{wang2025cega}: the adversary employs an iterative active learning strategy that adaptively selects informative nodes based on structural centrality, uncertainty, and diversity across multiple query cycles. We additionally replace its soft-label setting with hard labels. 4) DFEAII-Real \cite{DF_GNN_MSA_2024_USENIX}: since the original data-free variant suffers from prediction collapse, we follow the benchmarking protocol of \cite{zhao2026graphipbenchhardstealgraph} and instead query the victim on real induced subgraphs under the black-box hard-label setting. Please note that except for DFEAII-Real, MEA-Attack0 requires substantially higher query budgets than Dagger, whereas AdvMEA and CEGA rely on access to global information that are unavailable in practice.

\noindent\textit{\textbf{Implementations.}} Experiments are conducted on four Nvidia
A100 GPUs. Please see Appendix \ref{sec_experiments} for more details.

\noindent\textit{\textbf{Metrics.}} Except for accuracy and fidelity defined in Section \ref{ps}, we also consider 1) Macro F1 that evaluates all classes with equal weight, making it highly sensitive to performance degradation on minority classes; 2) Boundary fidelity that measures the 
prediction agreement between the surrogate and the victim on boundary nodes, defined as nodes where the victim's margin between its top two predicted probabilities 
falls below a threshold $\delta\!=\!0.2$. We empirically select $0.2$ as it achieves a better trade-off between the boundary space and number of samples located in this area. This metric captures the surrogate's ability to replicate the victim's decisions in the most challenging regions, where small perturbations can flip the predicted labels. 3) Number of queries, which is different from the number of accessible nodes. The adversary can adopt active learning or data synthesis and augmentation methods to reduce or expand the query node set. 

\subsection{Attack Performance (RQ1 and RQ2)}

\begin{table}[htbp]
\centering 
\caption{\centering The Performance of Dagger on Cora and Computer. Unit of Top 4 metrics: \textbf{1e-2}. Arrow indicates the direction of better performance and the \textbf{bold} font denotes the \textbf{`best'} results.}
\resizebox{\linewidth}{!}{
\begin{tabular}{c|c|c|ccccc}
\toprule
\multirow{2}{*}{\textbf{Dataset}}   & \multirow{2}{*}{\textbf{Backbone}} & \multirow{2}{*}{\textbf{Baseline}} & \multicolumn{5}{c}{\textbf{Metrics}}                                          \\ \cline{4-8}
                                    &                                    &                                    & ACC             & FID             & Macro F1              & Boundary Fid.           & Query \\ \hline
\multirow{24}{*}{\textbf{Cora}}     & \multirow{6}{*}{GCN}               & Rand                               & 26.57$\pm$0.00  & 27.31$\pm$0.00  & 6.00$\pm$0.00   & 10.00$\pm$0.00  & 135   \\
                                    &                                    & MEA                                & 61.50$\pm$2.63  & 64.33$\pm$2.30  & 53.08$\pm$2.37  & \textbf{37.62$\pm$4.10}  & 686   \\
                                    &                                    & AdvMEA                             & 40.22$\pm$3.95  & 42.19$\pm$5.77  & 28.47$\pm$6.43  & 19.05$\pm$7.59  & 315   \\
                                    &                                    & CEGA                               & 28.66$\pm$5.15  & 28.41$\pm$5.25  & 19.32$\pm$3.13  & 13.33$\pm$2.69  & 133   \\
                                    &                                    & DFEAII-Real                        & 29.64$\pm$3.87  & 29.64$\pm$4.11  & 13.87$\pm$3.51  & 10.48$\pm$0.67  & 135   \\
                                    &                                    & Dagger                             & \textbf{71.83$\pm$0.46}  & \textbf{73.68$\pm$0.17}  & \textbf{59.22$\pm$1.20}  & 32.86$\pm$1.17  & 135   \\ \cline{2-8}
                                    & \multirow{6}{*}{GAT}               & Rand                               & 28.78$\pm$3.13  & 29.52$\pm$3.65  & 8.43$\pm$3.31   & 11.11$\pm$3.93  & 135   \\
                                    &                                    & MEA                                & 59.66$\pm$2.44  & 62.85$\pm$3.22  & 51.99$\pm$2.97  & 31.94$\pm$1.96  & 686   \\
                                    &                                    & AdvMEA                             & 21.40$\pm$5.74  & 23.49$\pm$5.70  & 9.93$\pm$1.12   & 20.83$\pm$6.13  & 315   \\
                                    &                                    & CEGA                               & 36.29$\pm$7.36  & 38.25$\pm$8.45  & 21.88$\pm$8.14  & 22.22$\pm$9.67  & 133   \\
                                    &                                    & DFEAII-Real                        & 32.23$\pm$1.91  & 33.83$\pm$2.85  & 16.76$\pm$4.94  & 15.28$\pm$8.39  & 135   \\
                                    &                                    & Dagger                             & \textbf{74.28$\pm$1.25}  & \textbf{76.26$\pm$0.70}  & \textbf{67.08$\pm$1.25}  & \textbf{36.11$\pm$0.98}  & 135   \\ \cline{2-8}
                                    & \multirow{6}{*}{APPNP}             & Rand                               & 26.57$\pm$0.00  & 30.26$\pm$0.00  & 6.00$\pm$0.00   & 16.22$\pm$0.00  & 135   \\
                                    &                                    & MEA                                & 49.57$\pm$6.94  & 56.83$\pm$6.65  & 34.15$\pm$9.24  & 38.74$\pm$2.78  & 686   \\
                                    &                                    & AdvMEA                             & 27.31$\pm$1.09  & 31.00$\pm$1.09  & 7.64$\pm$1.47   & 15.77$\pm$0.64  & 315   \\
                                    &                                    & CEGA                               & 26.94$\pm$9.64  & 31.49$\pm$10.80 & 11.87$\pm$5.86  & 20.72$\pm$8.57  & 133   \\
                                    &                                    & DFEAII-Real                        & 31.73$\pm$5.52  & 35.18$\pm$6.51  & 15.93$\pm$3.29  & 18.47$\pm$2.30  & 135   \\
                                    &                                    & Dagger                             & \textbf{67.40$\pm$0.92}  & \textbf{74.29$\pm$1.06}  & \textbf{51.69$\pm$0.74}  & \textbf{41.89$\pm$1.91}  & 135   \\ \cline{2-8}
                                    & \multirow{6}{*}{GraphSAGE}         & Rand                               & 26.57$\pm$0.00  & 26.20$\pm$0.00  & 6.00$\pm$0.00   & 17.14$\pm$0.00  & 135   \\
                                    &                                    & MEA                                & 64.58$\pm$1.83  & 67.28$\pm$2.72  & 57.30$\pm$2.48  & \textbf{42.86$\pm$4.67}  & 686   \\
                                    &                                    & AdvMEA                             & 31.24$\pm$8.44  & 32.23$\pm$8.14  & 22.40$\pm$8.42  & 20.95$\pm$4.86  & 315   \\
                                    &                                    & CEGA                               & 45.26$\pm$2.26  & 45.26$\pm$2.05  & 38.87$\pm$5.55  & 26.67$\pm$11.74 & 133   \\
                                    &                                    & DFEAII-Real                        & 50.68$\pm$3.31  & 51.91$\pm$3.22  & 39.54$\pm$2.37  & 28.57$\pm$2.33  & 135   \\
                                    &                                    & Dagger                             & \textbf{76.51$\pm$0.92}  & \textbf{76.01$\pm$1.09}  & \textbf{71.91$\pm$1.53}  & 39.05$\pm$1.35  & 135   \\ \hline
\multirow{24}{*}{\textbf{Computer}} & \multirow{6}{*}{GCN}               & Rand                               & 41.13$\pm$0.00  & 43.31$\pm$0.00  & 5.83$\pm$0.00   & 28.21$\pm$0.00  & 687   \\
                                    &                                    & MEA                                & 76.24$\pm$1.61  & 82.66$\pm$1.81  & 47.60$\pm$3.91  & 47.86$\pm$5.37  & 8401  \\
                                    &                                    & AdvMEA                             & 62.52$\pm$4.39  & 67.61$\pm$5.36  & 43.50$\pm$6.25  & 42.74$\pm$7.72  & 480   \\
                                    &                                    & CEGA                               & 70.18$\pm$0.77  & 73.16$\pm$0.84  & \textbf{69.95$\pm$1.20}  & 42.74$\pm$6.97  & 690   \\
                                    &                                    & DFEAII-Real                        & 77.96$\pm$1.89  & 84.38$\pm$1.93  & 67.26$\pm$4.38  & 50.43$\pm$5.16  & 687   \\
                                    &                                    & Dagger                             & \textbf{81.73$\pm$0.71}  & \textbf{87.40$\pm$0.71}  & 67.52$\pm$1.42  & \textbf{60.26$\pm$3.63}  & 687   \\ \cline{2-8}
                                    & \multirow{6}{*}{GAT}               & Rand                               & 35.76$\pm$3.15  & 36.00$\pm$2.87  & 9.96$\pm$1.01   & 20.67$\pm$7.36  & 687   \\
                                    &                                    & MEA                                & 52.57$\pm$1.52  & 54.68$\pm$1.93  & 21.81$\pm$3.27  & 32.00$\pm$1.63  & 8401  \\
                                    &                                    & AdvMEA                             & 33.48$\pm$11.20 & 34.81$\pm$11.42 & 10.05$\pm$2.12  & 23.33$\pm$7.36  & 480   \\
                                    &                                    & CEGA                               & 54.29$\pm$5.74  & 55.60$\pm$5.45  & 40.11$\pm$6.37  & 26.67$\pm$3.77  & 690   \\
                                    &                                    & DFEAII-Real                        & 50.12$\pm$3.72  & 52.42$\pm$3.89  & 18.11$\pm$5.63  & \textbf{33.33$\pm$0.94}  & 687   \\
                                    &                                    & Dagger                             & \textbf{71.46$\pm$0.22}  & \textbf{72.84$\pm$0.24}  & \textbf{50.19$\pm$1.31}  & 31.33$\pm$0.94  & 687   \\ \cline{2-8}
                                    & \multirow{6}{*}{APPNP}             & Rand                               & 52.01$\pm$0.30  & 53.15$\pm$0.33  & 14.37$\pm$0.13  & 19.44$\pm$0.49  & 687   \\
                                    &                                    & MEA                                & 74.61$\pm$1.21  & 79.75$\pm$1.22  & 47.81$\pm$9.98  & 43.06$\pm$1.30  & 8401  \\
                                    &                                    & AdvMEA                             & 55.79$\pm$5.41  & 57.19$\pm$5.59  & 45.71$\pm$8.80  & 29.17$\pm$3.07  & 480   \\
                                    &                                    & CEGA                               & 51.09$\pm$12.42 & 50.85$\pm$12.91 & 51.65$\pm$10.61 & 28.82$\pm$4.28  & 690   \\
                                    &                                    & DFEAII-Real                        & 64.17$\pm$5.32  & 67.71$\pm$6.62  & 33.85$\pm$1.23  & 38.54$\pm$3.71  & 687   \\
                                    &                                    & Dagger                             & \textbf{84.33$\pm$0.36}  & \textbf{88.61$\pm$0.35}  & \textbf{77.75$\pm$2.04}  & \textbf{44.10$\pm$1.30}  & 687   \\ \cline{2-8}
                                    & \multirow{6}{*}{GraphSAGE}         & Rand                               & 41.13$\pm$0.00  & 41.86$\pm$0.00  & 5.83$\pm$0.00   & 27.42$\pm$0.00  & 687   \\
                                    &                                    & MEA                                & 76.11$\pm$3.13  & 80.01$\pm$3.62  & 59.54$\pm$5.41  & 45.70$\pm$3.31  & 8401  \\
                                    &                                    & AdvMEA                             & 48.96$\pm$20.16 & 49.78$\pm$21.37 & 40.31$\pm$19.11 & 33.87$\pm$3.95  & 480   \\
                                    &                                    & CEGA                               & 75.24$\pm$3.81  & 76.84$\pm$3.78  & \textbf{72.62$\pm$5.15}  & 38.71$\pm$4.75  & 690   \\
                                    &                                    & DFEAII-Real                        & 61.17$\pm$14.23 & 64.00$\pm$15.71 & 37.87$\pm$22.66 & 39.78$\pm$8.77  & 687   \\
                                    &                                    & Dagger                             & \textbf{80.62$\pm$0.33}  & \textbf{86.36$\pm$0.48}  & 72.36$\pm$1.21  & \textbf{47.31$\pm$2.01}  & 687    \\ \bottomrule
\end{tabular}}
\label{main_table}
\end{table}

\subsubsection{Effectiveness}
Due to space limitations, we put attack results of the rest two datasets in Appendix \ref{attack_app}. As shown in Table \ref{main_table} and \ref{main_table_app}, Dagger consistently outperforms all baselines in nearly every metric under the most stringent threat model, demonstrating the effectiveness of our framework. It even surpasses or achieves comparable attack performance against MEA, which usually submits queries 3.87-12.23$\times$ than Dagger. In terms of graph datasets, when using a similar number of queries, Dagger largely improves accuracy and fidelity over the SOTA baselines by up to 43.17\% and 45.27\%, respectively. It excels in graph datasets with much larger class spaces, like Cora and Computer. As for victim backbones, there is no single backbone inherently weaker or stronger against Dagger. When using APPNP as the victim backbone and trained on Cora, the query nodes cannot cover all classes, yielding relatively lower but still effective attack performance. When using GAT as the victim backbone trained on Computer, the query nodes are extremely imbalanced, causing all the baselines and Dagger to degrade compared to other victim backbones. These two anomalies are also reflected in macro F1 scores. Regarding boundary performance, Dagger also achieves outstanding boundary fidelity across most settings, indicating that the surrogate not only replicates the victim's confident predictions but also approximates its behaviors near decision boundaries. This is particularly relevant for downstream reconnaissance tasks such as crafting transferable adversarial examples \cite{IGNN_MSA_2022_SP}.

\subsubsection{Efficiency}
A key advantage of Dagger lies in its query efficiency. Dagger strictly requires 5\% nodes of the full graph and the number of queries is the exactly the same as the number of accessible nodes. MEA demands many more queries than all other baselines. Random generation and DFEAII-Real maintain the same query budgets as Dagger. The number of queries of AdvMEA and CEGA vary, AdvMEA possesses a larger query budget on Cora while a smaller budget than Dagger on the remaining graph datasets; CEGA owns the similar query budget to Dagger, it only increases queries compared to Dagger on Computer. Overall, Dagger is not only more attack-effective but also substantially more query-efficient. This efficiency is critical in practice, as a lower query footprint reduces the risk of detection by query-monitoring defenses \cite{prada}.

\subsection{Impact Factors (RQ3 to RQ5)} 
\subsubsection{GNN Architectures} 
Unlike existing stealing attacks, Dagger imposes no restrictions on requiring the same/similar GNN backbones or architecture configurations. Given that other graph datasets are relatively less prone to imbalance and class coverage issues while attaining compelling attack performance, we select Cora as the representative dataset for architecture sensitivity studies. In addition, GNNs typically employ shallow architectures to mitigate over-smoothing, a phenomenon where node representations become indistinguishable after multiple-round information propagation. Following standard settings \cite{gcn,gat, DF_GNN_MSA_2024_USENIX}, we set the number of layers as 2 or 3. As the computational complexity of GNN backbones scales linearly or quadratically with the hidden dimension and excessively large dimension might lead to overfitting problems especially on small-scale and sparse induced subgraphs obtained under limited query budgets, we evaluate the surrogate performance under compact hidden dimensions $d\in\{16,32,64\}$.

\begin{figure*}
    \centering
    \includegraphics[width=1\linewidth]{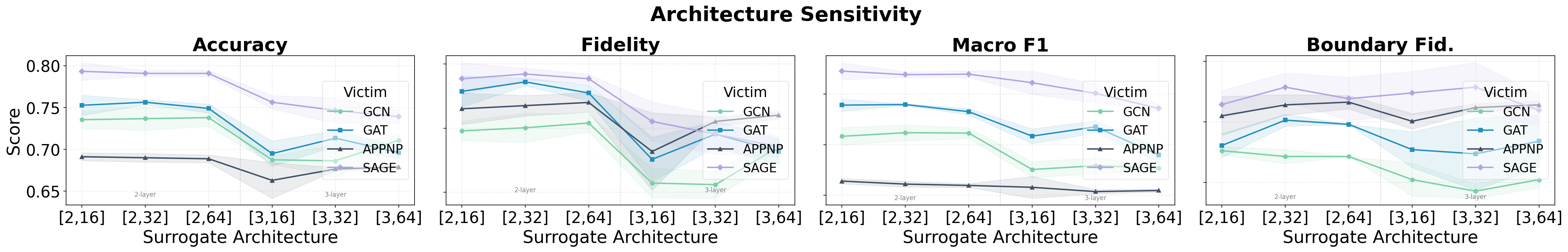}
    \caption{GNN Architecture Configuration Sensitivity}
    \label{gnn_backbone}
\end{figure*}
From Figure \ref{gnn_backbone}, Dagger remains stable to victim backbones and architecture configurations across all metrics, with performance variations remaining within approximately 1-4\%, confirming that the framework does not rely on a carefully tuned surrogate capacity to achieve competitive results. Comparing 2-layer and 3-layer configurations at the same hidden dimension, deeper architectures consistently perform on par with or slightly below their 2-layer counterparts. This suggests that additional depth does not provide meaningful benefit under the 5\% budget constraint. Increasing hidden dimension from 16 to 64 within the same depth also yields marginal changes, further indicating that the bottleneck lies in the quality and quantity of victim-predicted labels rather than model capacity. The default configuration $[2,16]$ achieves attack performance comparable to all larger architectures, even outperforming several specific victim backbones, such as GCN and APPNP. 

These empirical findings indicate a desirable attack property in practice: the adversary does not need prior knowledge of victim backbones or capacities, and a minimal architecture suffices to launch powerful GNN stealing attacks.

\subsubsection{Manifold}
\begin{figure*}
    \centering
    \begin{subfigure}[b]{0.99\linewidth}
        \centering
        \includegraphics[width=\linewidth]{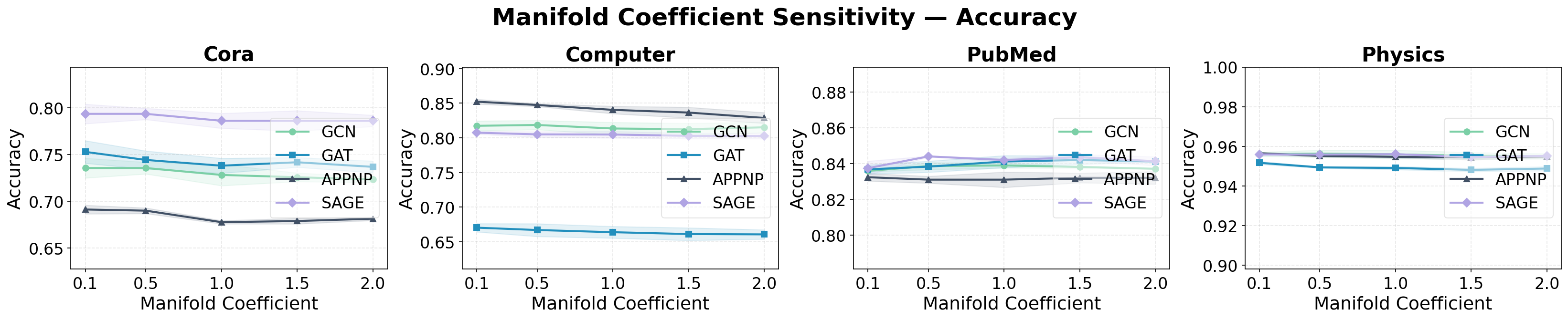}
        \label{fig:placeholder}
    \end{subfigure}

    \begin{subfigure}[b]{0.99\linewidth}
        \centering
        \includegraphics[width=\linewidth]{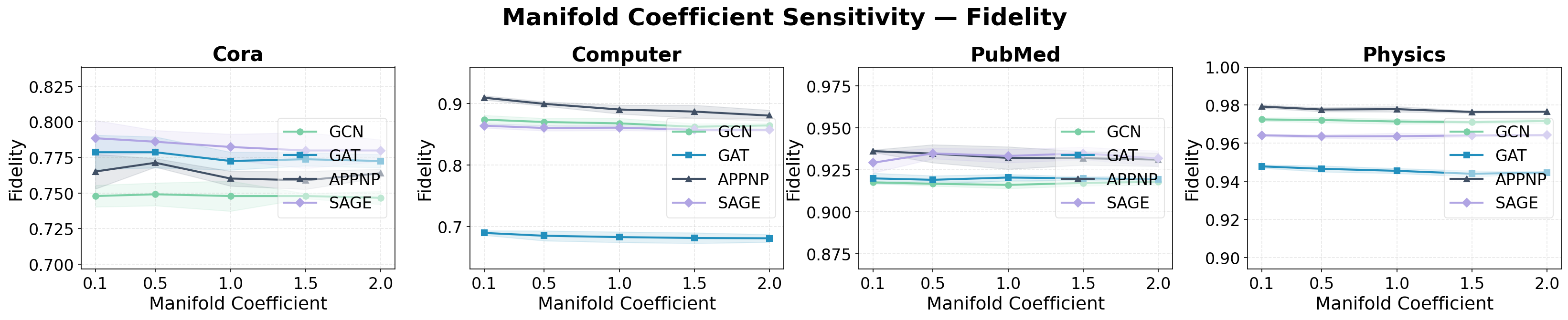}
        \label{fig:placeholder}
    \end{subfigure}

    \begin{subfigure}[b]{0.99\linewidth}
        \centering
        \includegraphics[width=\linewidth]{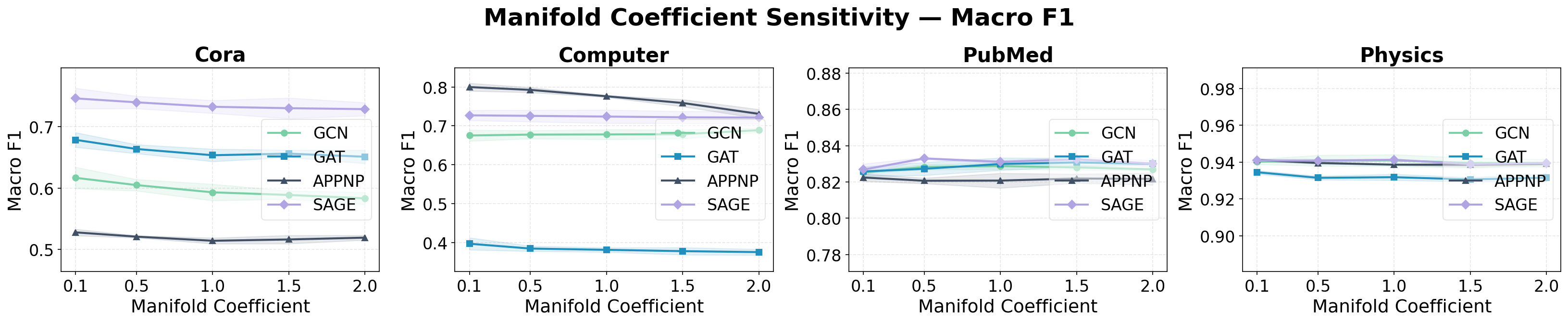}
        \label{fig:placeholder}
    \end{subfigure}

    \begin{subfigure}[b]{0.99\linewidth}
        \centering
        \includegraphics[width=\linewidth]{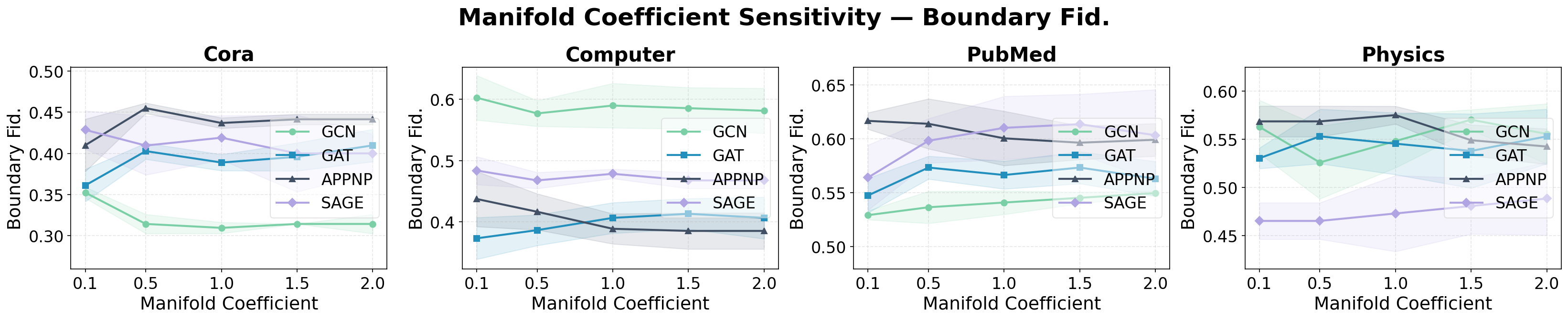}
        \label{fig:placeholder}
    \end{subfigure}
    \caption{Manifold Coefficient Sensitivity}
    \label{manifold}
\end{figure*}
As shown in Figure \ref{manifold}, Dagger demonstrates strong robustness to the manifold coefficient $w_{\text{mix}}$ across the range $[0.1,2.0]$, with performance fluctuations varying across graph datasets and victim backbones. On Physics and PubMed, performance curves are nearly flat across all victim backbones, indicating that the manifold regularization strength has negligible impact on larger and denser graphs with relatively smaller classes. On Cora, mild fluctuations are observed particularly in boundary fidelity, consistent with the fact that the induced subgraph is sparser and decision boundary regions are harder to approximate. Computer exhibits a moderate pattern, remaining largely stable with only a slight decline at larger coefficients. Consistent with previous findings, APPNP on Cora and GAT on Computer degrade due to class coverage and heavy imbalance issues. In all, there is no specific optimal manifold coefficient range, we accordingly tune this coefficient on different graph-backbone combinations. 

\subsubsection{Query Budget}

\begin{figure*}
    \centering
    \begin{subfigure}[b]{0.99\linewidth}
        \centering
        \includegraphics[width=\linewidth]{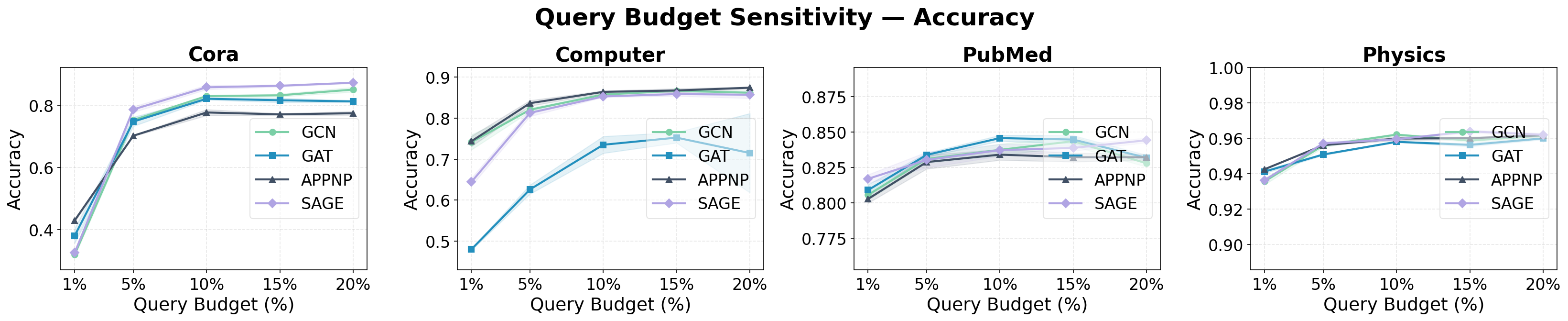}
        \label{fig:placeholder}
    \end{subfigure}

    \begin{subfigure}[b]{0.99\linewidth}
        \centering
        \includegraphics[width=\linewidth]{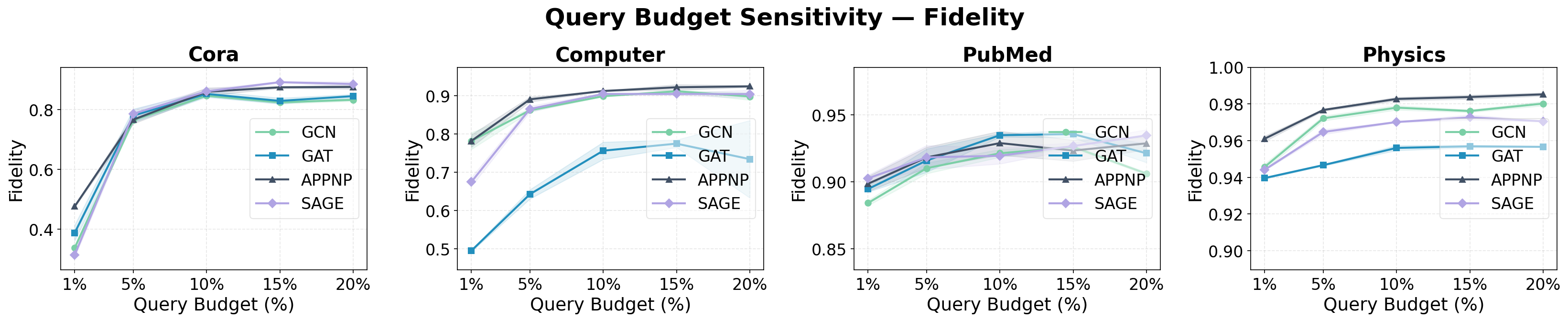}
        \label{fig:placeholder}
    \end{subfigure}

    \begin{subfigure}[b]{0.99\linewidth}
        \centering
        \includegraphics[width=\linewidth]{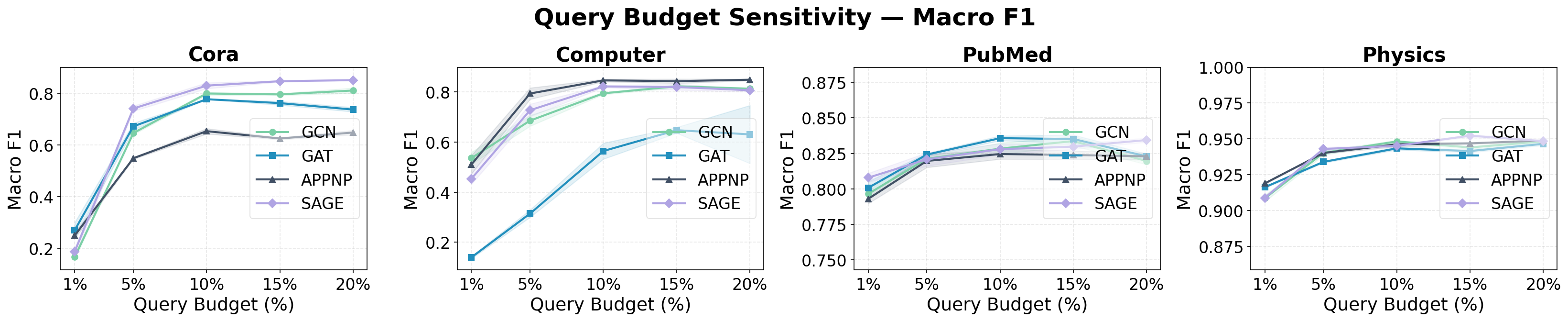}
        \label{fig:placeholder}
    \end{subfigure}

    \begin{subfigure}[b]{0.99\linewidth}
        \centering
        \includegraphics[width=\linewidth]{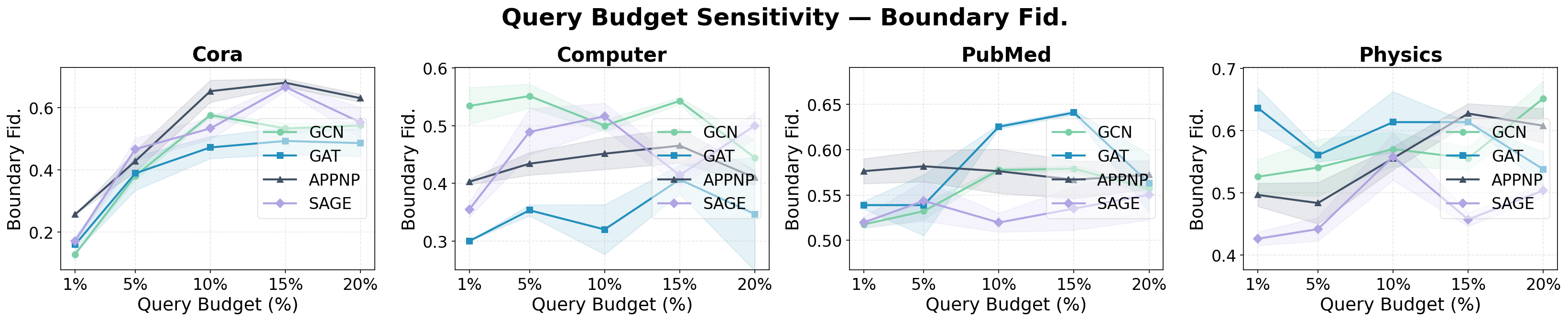}
        \label{fig:placeholder}
    \end{subfigure}
    \caption{Query Budget Sensitivity}
    \label{query_bug}
\end{figure*}

Figure \ref{query_bug} demonstrates that the attack performance of Dagger consistently improves as the query budget increases from 1\% to 20\% across all datasets and victim backbones, confirming that more query nodes provide richer supervision for the surrogate. Notably, Dagger already achieves competitive performance at the 5\% budget used in our main experiments, with diminishing returns beyond 10\% on most datasets. On Physics and PubMed, 1\% query budgets already yield competitive attack performance. Their accuracy, fidelity and F1 curves flatten around 5-10\% and remain stable through 20\%, suggesting that even a small fraction of training nodes is sufficient to capture the victim's inner behaviors on these datasets. On Cora and Computer, the improvement from 1\% to 5\% is more pronounced, reflecting the greater difficulty introduced by sparser induced subgraphs and more severe class imbalance at lower budgets. The boundary fidelity curves are more unstable than the other metrics, particularly on Computer, PubMed and Physics, we attribute this higher variance to the small number of boundary nodes in these datasets, where the victim model achieves higher overall confidence, leaving fewer nodes near the decision boundary and making the metric more susceptible to statistical fluctuation. Nevertheless, a clear upward trend is observed on Cora as budget increases. 

\subsubsection{Balance and Its Impact Factors} Due to space limitations, details are provided in Appendix \ref{sec_balance}.

\subsubsection{Ablation Studies}

\begin{table}[htbp]
\caption{\centering Ablation Studies across Graph Datasets and GNN backbones}
\centering
\resizebox{\linewidth}{!}{
\begin{tabular}{c|c|c|cccc}
\toprule
\multirow{2}{*}{\textbf{Dataset}}   & \multirow{2}{*}{\textbf{Backbone}} & \multirow{2}{*}{\textbf{Baseline}} & \multicolumn{4}{c}{\textbf{Metrics}}                              \\ \cline{4-7}
                                    &                                    &                                    & ACC            & FID            & Macro F1              & Boundary Fid.          \\ \hline
\multirow{16}{*}{\textbf{Cora}}     & \multirow{4}{*}{GCN}               & No manifold                        & 65.81$\pm$0.46 & 68.14$\pm$0.92 & 51.25$\pm$0.47 & 30.95$\pm$1.35 \\ 
                                    &                                    & No LA                              & 70.36$\pm$0.46 & 72.94$\pm$0.17 & 54.24$\pm$0.37 & 29.05$\pm$0.67 \\
                                    &                                    & No Tuning                         & 70.60$\pm$0.63 & 72.94$\pm$0.70 & 54.22$\pm$0.59 & 29.05$\pm$1.35 \\
                                    &                                    & Dagger                             & \textbf{71.83$\pm$0.46} & \textbf{73.68$\pm$0.17} & \textbf{59.22$\pm$1.20} & \textbf{32.86$\pm$1.17} \\ \cline{2-7}
                                    & \multirow{4}{*}{GAT}               & No manifold                        & 55.10$\pm$1.66 & 57.44$\pm$1.36 & 45.69$\pm$1.31 & 27.78$\pm$0.98 \\
                                    &                                    & No LA                              & 66.42$\pm$0.80 & 68.76$\pm$0.46 & 55.29$\pm$1.25 & 31.25$\pm$2.95 \\
                                    &                                    & No Tuning                         & 60.76$\pm$2.22 & 62.85$\pm$2.63 & 51.34$\pm$2.48 & 30.56$\pm$1.96 \\
                                    &                                    & Dagger                             & \textbf{74.28$\pm$1.25} & \textbf{76.26$\pm$0.70} & \textbf{67.08$\pm$1.25} & \textbf{36.11$\pm$0.98} \\
                                    \cline{2-7}
                                    & \multirow{4}{*}{APPNP}             & No manifold                        & 63.96$\pm$0.35 & 71.59$\pm$0.30 & 48.09$\pm$0.54 & 40.09$\pm$2.55 \\
                                    &                                    & No LA                              & 65.44$\pm$0.70 & 73.06$\pm$1.09 & 48.85$\pm$0.82 & 40.99$\pm$2.30 \\
                                    &                                    & No Tuning                         & 62.61$\pm$0.87 & 68.02$\pm$0.46 & 46.89$\pm$0.71 & 32.88$\pm$3.37 \\
                                    &                                    & Dagger                             & \textbf{67.40$\pm$0.92} & \textbf{74.29$\pm$1.06} & \textbf{51.69$\pm$0.74} & \textbf{41.89$\pm$1.91} \\
                                    \cline{2-7}
                                    & \multirow{4}{*}{GraphSAGE}         & No manifold                        & 62.85$\pm$0.17 & 62.36$\pm$0.30 & 52.12$\pm$0.04 & 25.71$\pm$0.00 \\
                                    &                                    & No LA                              & 70.85$\pm$0.30 & 69.50$\pm$0.17 & 61.77$\pm$0.61 & 32.38$\pm$3.56 \\
                                    &                                    & No Tuning                         & 64.70$\pm$1.71 & 63.59$\pm$1.66 & 53.07$\pm$1.66 & 28.57$\pm$2.33 \\
                                    &                                    & Dagger                             & \textbf{76.51$\pm$0.92} & \textbf{76.01$\pm$1.09} & \textbf{71.91$\pm$1.53} & \textbf{39.05$\pm$1.35} \\ \hline

        \multirow{16}{*}{\textbf{Computer}} & \multirow{4}{*}{GCN}               & No manifold                        & 80.35$\pm$0.36 & 85.51$\pm$0.55 & 62.33$\pm$1.67 & 53.85$\pm$2.77 \\
                                    &                                    & No LA                              & 80.86$\pm$0.72 & 86.63$\pm$0.77 & 65.38$\pm$0.81 & 55.13$\pm$3.63 \\
                                    &                                    & No Tuning                         & 80.21$\pm$0.36 & 85.59$\pm$0.50 & 64.43$\pm$1.48 & 57.26$\pm$3.36 \\
                                    &                                    & Dagger                             & \textbf{81.73$\pm$0.71} & \textbf{87.40$\pm$0.71} & \textbf{67.52$\pm$1.42} & \textbf{60.26$\pm$3.63} \\
                                    \cline{2-7}
                                    & \multirow{4}{*}{GAT}               & No manifold                        & 64.97$\pm$0.27 & 66.72$\pm$0.54 & 34.80$\pm$0.74 & 31.33$\pm$0.94 \\
                                    &                                    & No LA                              & 67.30$\pm$0.33 & 68.48$\pm$0.69 & 39.62$\pm$0.52 & 30.00$\pm$1.63 \\
                                    &                                    & No Tuning                         & 65.60$\pm$0.27 & 67.42$\pm$0.38 & 35.61$\pm$1.02 & \textbf{32.67$\pm$0.94} \\
                                    &                                    & Dagger                             & \textbf{71.46$\pm$0.22} & \textbf{72.84$\pm$0.24} & \textbf{50.19$\pm$1.31} & 31.33$\pm$0.94 \\
                                    \cline{2-7}
                                    & \multirow{4}{*}{APPNP}             & No manifold                        & 81.32$\pm$0.33 & 86.36$\pm$0.33 & 62.53$\pm$2.61 & 43.75$\pm$1.70 \\
                                    &                                    & No LA                              & 83.33$\pm$0.45 & 88.03$\pm$0.24 & 73.37$\pm$0.66 & \textbf{44.44$\pm$2.14} \\
                                    &                                    & No Tuning                         & 82.82$\pm$0.39 & 87.96$\pm$0.28 & 71.24$\pm$0.33 & 43.06$\pm$0.98 \\
                                    &                                    & Dagger                             & \textbf{84.33$\pm$0.36} & \textbf{88.61$\pm$0.35} & \textbf{77.75$\pm$2.04} & 44.10$\pm$1.30 \\
                                    \cline{2-7}
                                    & \multirow{4}{*}{GraphSAGE}         & No manifold                        & 77.62$\pm$1.23 & 82.80$\pm$1.47 & 62.29$\pm$3.05 & 47.31$\pm$2.74 \\
                                    &                                    & No LA                              & 79.17$\pm$0.42 & 84.84$\pm$0.42 & 68.48$\pm$0.52 & 47.31$\pm$3.31 \\
                                    &                                    & No Tuning                         & 77.18$\pm$0.62 & 82.68$\pm$0.91 & 64.01$\pm$0.83 & \textbf{47.85$\pm$1.52} \\
                                    &                                    & Dagger                             & \textbf{80.62$\pm$0.33} & \textbf{86.36$\pm$0.48} & \textbf{72.36$\pm$1.21} & 47.31$\pm$2.01 \\ \bottomrule
\end{tabular}}
\label{ablation_1}
\end{table}
To systematically assess the contribution of each component in Dagger, we conduct a series of ablation studies across graph datasets and victim backbones. The ablation results confirm the effectiveness of each component in Dagger, and the full model achieves the best or near-best performance in the majority of graph-backbone combinations. 

In the Decoupling-based Surrogate Training phase, since the knowledge distillation loss is indispensable for model stealing attacks, we instead remove the manifold mixup regularization module (No Manifold). In the Imbalance-aware Classifier Tuning phase, where rebalancing is essential, we retain the class-balanced sampling and remove logit adjustment (No LA). We also evaluate a variant that retains only Phase 1 while omitting Phase 2 (No Tuning). The ablation results are presented in Table
\ref{ablation_1} and \ref{ablation_2}.

No Manifold consistently degrades performance across Cora and Computer, with the most noticeable drops observed in GAT and GraphSAGE. The degradation on PubMed and Physics is less pronounced, where the larger and relatively balanced query set, denser graph structure, and smaller class space already provide sufficient supervision signals, making the additional mixup regularization less critical. 

No LA leads to persistent performance drops, particularly on Cora and Computer, where class imbalance is most severe under the strict 5\% budget. On PubMed and Physics, where the class distribution is more balanced and the victim exhibits higher confidence, the contribution of LA is smaller yet still measurable. Consistent with prior findings, when the query set distribution is balanced, enforcing LA may introduce overfitting and cause slight performance drops.

Similar to No LA, No Tuning still results in substantial performance degradation across all victim backbones on Cora and Computer while yielding a slight decline on PubMed and Physics. Under severe imbalance, No Tuning consistently performs worse than No LA, implying that class-balanced sampling plays a fundamental role in stabilizing classifier tuning under skewed query distributions.

\subsection{Defense (RQ6)}
To evaluate Dagger's robustness under SOTA defense mechanisms, we adopt two distinct defenses, namely, PRADA \cite{prada} and BackdoorWM \cite{backdoorwm}. PRADA monitors the statistical distribution of consecutive queries to identify deviations from a normal distribution, achieving a claimed perfect detection rate against model stealing attacks. Its core assumption is that legitimate queries follow a natural data distribution, whereas model stealing attacks generate statistically anomalous query patterns. BackdoorWM trains a watermarked GNN model to enable post-hoc black-box ownership verification. Specifically, for node classification tasks, a fixed trigger pattern is injected into a small subset of training nodes associated with a target label. Any surrogate inheriting the victim's decision boundaries is expected to replicate this backdoor behavior on triggered inputs, facilitating near-perfect ownership verification.

We follow the default parameter setting in defense benchmarking \cite{zhao2026graphipbenchhardstealgraph} and report their defense results in Table \ref{defense_table}. We observe that Dagger remains highly effective against both PRADA and BackdoorWM across all graph datasets and victim backbones, achieving attack performance comparable to the undefended setting. Under most graph-backbone combinations, Dagger evaluated against BackdoorWM (short for BackdoorWM, hereinafter) outperforms the one against PRADA (short for PRADA, hereinafter), which aligns with imbalance trends reflected by macro F1. Notably,  on challenging graphs such as Cora and Computer, BackdoorWM even obtains better attack performance than the undefended setting, this variation can be attributed to subtle differences between the clean and the backdoor-injected victims, as the backdoor injection slightly alters the victim's prediction distribution. 

PRADA is ineffective. PRADA achieves 100\% prediction fidelity relative to the original victim model, suggesting that it fails to trigger an alarm and treats every query as legitimate during initial model evaluation. Although PRADA eventually activates in the subsequent querying phase and corrupts a portion of the victim-predicted labels with Gaussian noise, Dagger's attack performance remains comparable to the undefended setting. One plausible explanation is that the trigger occurs in the late query phase, Dagger has already accumulated sufficient clean supervision signals from early queries, and its manifold mixup confers inherent robustness against label noise by interpolating latent representations.

BackdoorWM provides negligible protection. As for BackdoorWM, fidelity drops to between 76.75\% and 99.25\% across datasets and backbones, indicating that the watermark is only partially inherited by the surrogate. More importantly, BackdoorWM yields slightly higher attack performance than or comparable to the undefended setting across all configurations, confirming that BackdoorWM fails to degrade the functional quality of the stolen surrogate model.



\begin{table}[htbp]
\caption{\centering Dagger's Attack Performance against Defense Mechanisms across Graph Datasets and GNN Backbones}
\resizebox{\linewidth}{!}{
\begin{tabular}{c|c|c|cc|cccc}
\toprule
\multirow{2}{*}{\textbf{Dataset}} & \multirow{2}{*}{\textbf{Backbone}} & \multirow{2}{*}{\textbf{Defense}} & \multirow{2}{*}{\textbf{Pred Acc}} & \multirow{2}{*}{\textbf{Pred Fid}} & \multicolumn{4}{c}{\textbf{Metrics}}                              \\ \cline{6-9}
                                  &                                    &                                    &                                    &                                    & ACC            & FID            & Macro F1              & Boundary Fid.          \\ \hline
\multirow{8}{*}{\textbf{Cora}}             & \multirow{2}{*}{GCN}               & PRADA                              & 83.03                              & 100.00                             & 68.88$\pm$0.17 & 72.45$\pm$0.63 & 55.80$\pm$0.64 & 33.33$\pm$1.78 \\
                                  &                                    & BackdoorWM                         & 85.98                              & 84.87                              & 75.15$\pm$1.36 & 75.77$\pm$0.97 & 69.79$\pm$1.42 & 41.43$\pm$1.17 \\ \cline{2-9}
                                  & \multirow{2}{*}{GAT}               & PRADA                              & 80.81                              & 100.00                             & 72.69$\pm$1.38 & 74.91$\pm$0.60 & 65.61$\pm$2.25 & 40.97$\pm$2.60 \\
                                  &                                    & BackdoorWM                         & 85.98                              & 84.87                              & 74.66$\pm$0.97 & 76.01$\pm$1.68 & 68.96$\pm$1.08 & 41.67$\pm$5.10 \\
                                  \cline{2-9}
                                  & \multirow{2}{*}{APPNP}             & PRADA                              & 80.44                              & 100.00                             & 68.02$\pm$0.46 & 73.55$\pm$0.35 & 53.95$\pm$0.63 & 45.95$\pm$1.10 \\
                                  &                                    & BackdoorWM                         & 88.19                              & 76.01                              & 77.98$\pm$0.46 & 76.63$\pm$0.63 & 73.28$\pm$0.70 & 49.10$\pm$1.69 \\
                                  \cline{2-9}
                                  & \multirow{2}{*}{GraphSAGE}         & PRADA                              & 87.45                              & 100.00                             & 69.13$\pm$0.35 & 69.74$\pm$0.80 & 61.92$\pm$0.16 & 26.67$\pm$1.35 \\
                                  &                                    & BackdoorWM                         & 89.30                              & 92.25                              & 72.82$\pm$0.46 & 72.69$\pm$0.52 & 69.38$\pm$0.51 & 39.05$\pm$1.35 \\ \hline
\multirow{8}{*}{\textbf{PubMed}}           & \multirow{2}{*}{GCN}               & PRADA                              & 84.28                              & 100.00                             & 83.81$\pm$0.10 & 91.60$\pm$0.20 & 82.75$\pm$0.13 & 55.26$\pm$1.64 \\
                                  &                                    & BackdoorWM                         & 85.09                              & 96.60                              & 84.58$\pm$0.31 & 91.85$\pm$0.05 & 83.54$\pm$0.31 & 55.12$\pm$0.55 \\
                                  \cline{2-9}
                                  & \multirow{2}{*}{GAT}               & PRADA                              & 84.84                              & 100.00                             & 83.62$\pm$0.18 & 92.00$\pm$0.29 & 82.53$\pm$0.20 & 54.92$\pm$1.47 \\
                                  &                                    & BackdoorWM                         & 85.04                              & 97.77                              & 83.92$\pm$0.25 & 92.22$\pm$0.35 & 82.87$\pm$0.17 & 56.82$\pm$0.65 \\
                                  \cline{2-9}
                                  & \multirow{2}{*}{APPNP}             & PRADA                              & 85.60                              & 100.00                             & 83.23$\pm$0.21 & 93.61$\pm$0.07 & 82.24$\pm$0.21 & 61.67$\pm$0.76 \\
                                  &                                    & BackdoorWM                         & 87.27                              & 95.28                              & 84.62$\pm$0.17 & 94.81$\pm$0.25 & 83.63$\pm$0.17 & 67.48$\pm$1.94 \\
                                  \cline{2-9}
                                  & \multirow{2}{*}{GraphSAGE}         & PRADA                              & 86.00                              & 100.00                             & 83.76$\pm$0.38 & 92.93$\pm$0.58 & 82.70$\pm$0.29 & 56.41$\pm$3.02 \\
                                  &                                    & BackdoorWM                         & 86.82                              & 95.99                              & 83.77$\pm$0.26 & 92.21$\pm$0.51 & 82.75$\pm$0.25 & 52.82$\pm$2.33 \\ \hline
\multirow{8}{*}{\textbf{Computer}}         & \multirow{2}{*}{GCN}               & PRADA                              & 89.97                              & 100.00                             & 81.61$\pm$0.66 & 87.48$\pm$0.81 & 66.88$\pm$1.77 & 59.40$\pm$3.02 \\
                                  &                                    & BackdoorWM                         & 89.03                              & 95.20                              & 80.21$\pm$0.73 & 86.07$\pm$0.39 & 71.69$\pm$1.46 & 53.85$\pm$1.81 \\
                                  \cline{2-9}
                                  & \multirow{2}{*}{GAT}               & PRADA                              & 88.59                              & 100.00                             & 69.57$\pm$0.45 & 70.59$\pm$0.29 & 45.04$\pm$0.68 & 32.67$\pm$2.49 \\
                                  &                                    & BackdoorWM                         & 90.41                              & 92.30                              & 77.59$\pm$0.30 & 77.28$\pm$0.19 & 59.76$\pm$1.22 & 44.00$\pm$4.32 \\
                                  \cline{2-9}
                                  & \multirow{2}{*}{APPNP}             & PRADA                              & 89.39                              & 100.00                             & 84.23$\pm$0.21 & 88.35$\pm$0.30 & 77.75$\pm$1.49 & 40.62$\pm$2.55 \\
&                                    & BackdoorWM                         & 88.74                              & 93.02                              & 81.73$\pm$0.55 & 83.50$\pm$0.45 & 71.35$\pm$2.51 & 36.46$\pm$1.70 \\
                                  \cline{2-9}
                                  & \multirow{2}{*}{GraphSAGE}         & PRADA                              & 90.12                              & 100.00                             & 80.77$\pm$0.19 & 86.43$\pm$0.35 & 72.71$\pm$1.32 & 48.39$\pm$2.28 \\
                                  &                                    & BackdoorWM                         & 88.81                              & 94.84                              & 78.75$\pm$0.65 & 82.95$\pm$0.06 & 70.09$\pm$1.63 & 37.10$\pm$1.32 \\ \hline
\multirow{8}{*}{\textbf{Physics}}          & \multirow{2}{*}{GCN}               & PRADA                              & 96.26                              & 100.00                             & 95.60$\pm$0.17 & 97.18$\pm$0.03 & 94.07$\pm$0.27 & 56.30$\pm$1.05 \\
                                  &                                    & BackdoorWM                         & 95.91                              & 97.86                              & 95.20$\pm$0.07 & 96.77$\pm$0.05 & 93.58$\pm$0.07 & 50.37$\pm$2.10 \\
                                  \cline{2-9}
                                  & \multirow{2}{*}{GAT}               & PRADA                              & 94.99                              & 100.00                             & 95.16$\pm$0.04 & 94.82$\pm$0.08 & 93.42$\pm$0.07 & 54.55$\pm$1.86 \\
                                  &                                    & BackdoorWM                         & 95.22                              & 95.42                              & 95.36$\pm$0.09 & 94.71$\pm$0.04 & 93.57$\pm$0.09 & 55.30$\pm$2.14 \\
                                  \cline{2-9}
                                  & \multirow{2}{*}{APPNP}             & PRADA                              & 96.70                              & 100.00                             & 95.66$\pm$0.08 & 97.92$\pm$0.07 & 94.13$\pm$0.10 & 56.86$\pm$2.77 \\
                                  &                                    & BackdoorWM                         & 96.81                              & 98.84                              & 95.44$\pm$0.08 & 97.17$\pm$0.13 & 93.94$\pm$0.02 & 54.25$\pm$0.92 \\
                                  \cline{2-9}
                                  & \multirow{2}{*}{GraphSAGE}         & PRADA                              & 95.83                              & 100.00                             & 95.65$\pm$0.09 & 96.45$\pm$0.05 & 94.18$\pm$0.12 & 45.74$\pm$2.90 \\
                                  &                                    & BackdoorWM                         & 95.71                              & 99.25                              & 95.36$\pm$0.04 & 96.65$\pm$0.10 & 93.65$\pm$0.06 & 44.96$\pm$2.19 \\ \bottomrule
\end{tabular}}
\label{defense_table}
\end{table}

\noindent\textbf{Limitations and Future Work.} First, our analysis reveals that extreme minority classes with near-zero class coverage represent a fundamental limitation, motivating the development of adaptive query strategies that allocate budget more effectively. Second, despite operating under a strict query budget, further reducing the number of queries through active learning methods remains a key priority. Third, while we focus on mainstream message-passing GNNs that implicitly assume homophily, tailoring model stealing attacks to heterogeneous graphs and GNNs represents a promising future direction. Finally, designing robust defenses against Dagger under such highly constrained threat model represents a critical open problem for the research community.

\section{Conclusion}
\label{sec:conclusion}

In this work, we propose Dagger, a novel model stealing attack against GNNs under a strictly constrained black-box, hard-label, query-limited, and backbone-agnostic threat model. We identify four fundamental challenges specific to this setting: poor surrogate performance, insufficient supervision, class imbalance with class coverage issues, and backbone mismatch. Motivated by these systematic limitations, we design a two-phase decoupling-based attack framework. The first phase leverages decoupled information diffusion to incorporate isolated nodes and mitigate structural sparsity, and manifold-level node mixup to compensate for the incomplete and imbalanced supervision signals. The second phase applies a decoupled classifier fine-tuned with class-balanced sampling and logit adjustment to recover minority-class performance without requiring extra victim queries. Extensive experiments demonstrate that Dagger consistently outperforms state-of-the-art baselines across diverse graph datasets and victim backbones. Moreover, our defense evaluations reveal that Dagger is inherently resistant to both query-monitoring and backdoor watermarking defenses, highlighting the urgent need for more robust GNN defense mechanisms.

\section*{Open Science}
To comply with the CFP’s Open Science guidelines, we release the source code along with implementation details and evaluation pipeline to reproduce our experiments and conduct the attacks. As the datasets used in our paper are all downloadable from Pytorch Geometric \cite{pyg}, we direct users to the downloading links and provide processing scripts in our artifacts. To preserve anonymity during review, these materials are hosted in an anonymous repository \url{https://anonymous.4open.science/r/Dagger-1D14}, and will be migrated to a public GitHub repository upon acceptance. This artifact release is intended to facilitate more in-depth adversarial investigation on graph neural networks and to motivate future work on more robust and effective defense mechanisms.

\section*{LLM Usage Considerations}
We use Claude only for editorial assistance in this manuscript, and all outputs were inspected by the authors to ensure accuracy and originality.

\section*{Ethical Considerations}
This study investigates the vulnerability of graph neural networks (GNNs) through the lens of model stealing attacks and the empirical results demonstrate GNNs are highly susceptible to such attacks even under a more restricted threat model than existing GNN stealing attacks. 

We recognize the inherent dual-use risk in publishing attack methodologies. However, we contend that the benefits to the research community outweigh this risk. First, the threat model we explore reflects realistic deployments that existing defenses have not adequately addressed, and our findings underscore the practical severity of GNN stealing attacks and motivate the urgent development of corresponding defense mechanisms. Second, the primary beneficiaries of this work are defenders. By characterizing the attack surface, we enable model developers and deployment teams to proactively harden their systems before adversarial exploitation occurs in the wild. Third, as a red team, we conduct all experiments on local devices using publicly available graph datasets. The testing environment does not interact with real-world third-party systems, collect personal data, or involve human subjects. Finally, our public artifact release is scoped to the code and scripts necessary to reproduce the reported results. We do not release components that would trivially enable large-scale exploitation beyond the experimental setup described in this paper.



\bibliographystyle{IEEEtranS}
\bibliography{Reference}


\appendix




\subsection{Implementation}
\label{sec_experiments}
We follow the standard training/validation/test splits of 80\%/10\%/10\% \cite{hu2020pretraining, lin2020graphneuralnetworksincluding, break_limit} across four graph datasets. As for victim GNN training, we adopt the learning rate as $0.001$, weight decay as $5e\!-\!4$ and training epochs as $600$. Adam is used as the optimizer. The hidden dimension is $16$ and the number of layers is $2$. In terms of GNN stealing attacks, we choose the learning rate from $\{0.01, 0.001\}$, weight decay as $5e\!-\!4$ and training epochs as $600$. $K\in\{5,10,15,20\}$, $\alpha\in\{0.1, 0.15, 0.2,0.25,0.3\}$ and $w_{\text{mix}}\in\{0.1,0.5,1,1.5\}$ for surrogate pre-training; We set the learning rate as $0.01$, training epochs as $200$ and $\tau\!=\!0.5$ for imbalance-aware fine-tuning. Regarding baselines, we leverage the same training configurations for fair comparison and other customized parameters strictly follow the default setting in \cite{zhao2026graphipbenchhardstealgraph}. We further restrict node selection to training node sets for two reasons, first, test nodes are inaccessible to the adversary in realistic deployments; second, covering test nodes in the query set would introduce data leakage into the evaluation, artificially inflating the attack performance of the surrogate model. Notably, even under these restrictions, the baselines operate under more favorable conditions than Dagger as their victim-surrogate backbones are aligned while Dagger is backbone-agnostic. We finally report the average results of $3$ trials. All experiments are conducted on a $64$-bit machine with $4$ Nvidia A$100$ GPUs. 

\subsection{Benign GNN Performance}
\label{sec_benign}

Table \ref{benign_gnn} reports the benign classification performance of each victim GNN backbone. All backbones achieve strong predictive performance across graph datasets, confirming that they represent valuable stealing targets. Among these, GraphSAGE generally outperforms other backbones on Cora, PubMed and Computer, while APPNP excels on Physics.  

\begin{table}[htbp]
\caption{Benign Performance across GNN Backbones}
\centering
\begin{tabular}{c|c|cc}
\toprule
\multirow{2}{*}{\textbf{Dataset}}  & \multicolumn{1}{c|}{\multirow{2}{*}{\textbf{GNN}}} & \multicolumn{2}{c}{\textbf{Metrics}} \\ \cline{3-4} 
                          & \multicolumn{1}{c|}{}                     & Acc          & F1           \\ \hline
\multirow{4}{*}{\textbf{Cora}}     & GCN                                      & 83.03        & 79.61        \\
                          & GAT                                      & 80.81        & 73.26        \\
                          & APPNP                                    & 80.44        & 67.30        \\
                          & GraphSAGE                                & \textbf{87.45}        & \textbf{85.48}        \\ \hline
\multirow{4}{*}{\textbf{PubMed}}   & GCN                                      & 84.28        & 83.18        \\
                          & GAT                                      & 84.84        & 83.63        \\
                          & APPNP                                    & \textbf{86.00}        & 84.44        \\
                          & GraphSAGE                                & \textbf{86.00}        & \textbf{85.01}        \\ \hline
\multirow{4}{*}{\textbf{Computer}} & GCN                                      & 89.97        & 88.64        \\
                          & GAT                                      & 88.59        & 86.88        \\
                          & APPNP                                    & 89.39        & 86.94        \\
                          & GraphSAGE                                & \textbf{90.12}        & \textbf{88.76}        \\ \hline
\multirow{4}{*}{\textbf{Physics}}  & GCN                                      & 96.26        & 95.00        \\
                          & GAT                                      & 94.99        & 93.34        \\
                          & APPNP                                    & \textbf{97.00}        & \textbf{95.72}        \\
                          & GraphSAGE                                & 95.83        & 94.48
                          \\ \bottomrule
\end{tabular}
\label{benign_gnn}
\end{table}

\subsection{Attack Performance}
\label{attack_app}
\begin{table}[htbp]
\centering 
\caption{\centering The Performance of Dagger on PubMed and Physics. Unit: \textbf{1e-2}. Arrow indicates the direction of better performance and the \textbf{bold} font denotes the \textbf{`best'} results.}
\resizebox{\linewidth}{!}{
\begin{tabular}{c|c|c|ccccc}
\toprule
\multirow{2}{*}{\textbf{Dataset}}   & \multirow{2}{*}{\textbf{Backbone}} & \multirow{2}{*}{\textbf{Baseline}} & \multicolumn{5}{c}{\textbf{Metrics}}                                          \\ \cline{4-8}
                                    &                                    &                                    & ACC             & FID             & Macro F1              & Boundary Fid.           & Query \\ \hline
\multirow{24}{*}{\textbf{PubMed}}   & \multirow{6}{*}{GCN}               & Rand                               & 40.67$\pm$0.00  & 39.15$\pm$0.00  & 19.27$\pm$0.00  & 39.04$\pm$0.00  & 985   \\
                                    &                                    & MEA                                & 82.44$\pm$0.57  & \textbf{93.93$\pm$1.48}  & 81.26$\pm$0.65  & \textbf{63.74$\pm$1.03}  & 3810  \\
                                    &                                    & AdvMEA                             & 50.91$\pm$5.15  & 53.23$\pm$5.90  & 44.38$\pm$10.17 & 51.52$\pm$3.48  & 150   \\
                                    &                                    & CEGA                               & 50.68$\pm$1.21  & 54.36$\pm$1.18  & 47.83$\pm$2.72  & 42.54$\pm$0.62  & 984   \\
                                    &                                    & DFEAII-Real                        & 49.15$\pm$2.47  & 52.18$\pm$2.35  & 42.51$\pm$3.44  & 39.04$\pm$1.29  & 985   \\
                                    &                                    & Dagger                             & \textbf{83.81$\pm$0.06}  & 91.72$\pm$0.17  & \textbf{82.80$\pm$0.14}  & 54.53$\pm$0.55  & 985   \\ \cline{2-8}
                                    & \multirow{6}{*}{GAT}               & Rand                               & 40.67$\pm$0.00  & 38.74$\pm$0.00  & 19.27$\pm$0.00  & 39.90$\pm$0.00  & 985   \\
                                    &                                    & MEA                                & 82.49$\pm$0.31  & \textbf{94.32$\pm$0.53}  & 81.33$\pm$0.42  & \textbf{60.28$\pm$3.12}  & 3810  \\
                                    &                                    & AdvMEA                             & 33.67$\pm$9.90  & 32.59$\pm$8.70  & 16.50$\pm$3.92  & 34.54$\pm$7.57  & 150   \\
                                    &                                    & CEGA                               & 76.17$\pm$2.29  & 84.99$\pm$2.12  & 75.48$\pm$2.12  & 50.43$\pm$2.33  & 984   \\
                                    &                                    & DFEAII-Real                        & 78.06$\pm$0.77  & 86.76$\pm$1.90  & 76.32$\pm$1.78  & 50.43$\pm$3.79  & 985   \\
                                    &                                    & Dagger                             & \textbf{83.54$\pm$0.32}  & 92.29$\pm$0.27  & \textbf{82.41$\pm$0.33}  & 56.65$\pm$1.29  & 985   \\ \cline{2-8}
                                    & \multirow{6}{*}{APPNP}             & Rand                               & 50.19$\pm$13.46 & 51.20$\pm$16.76 & 30.09215.30     & 45.48$\pm$4.20  & 985   \\
                                    &                                    & MEA                                & 81.36$\pm$0.91  & 89.54$\pm$1.65  & 79.37$\pm$0.83  & \textbf{62.62$\pm$1.91}  & 3810  \\
                                    &                                    & AdvMEA                             & 41.84$\pm$0.90  & 42.49$\pm$1.29  & 24.18$\pm$2.59  & 38.60$\pm$2.67  & 150   \\
                                    &                                    & CEGA                               & 60.09$\pm$8.94  & 62.90$\pm$9.64  & 59.45$\pm$8.48  & 41.70$\pm$2.01  & 984   \\
                                    &                                    & DFEAII-Real                        & 64.37$\pm$4.60  & 67.44$\pm$6.38  & 54.58$\pm$9.96  & 49.12$\pm$0.83  & 985   \\
                                    &                                    & Dagger                             & \textbf{83.27$\pm$0.15}  & \textbf{93.68$\pm$0.13}  & \textbf{82.29$\pm$0.14}  & 62.48$\pm$1.16  & 985   \\ \cline{2-8}
                                    & \multirow{6}{*}{GraphSAGE}         & Rand                               & 40.67$\pm$0.00  & 39.86$\pm$0.00  & 19.27$\pm$0.00  & 48.72$\pm$0.00  & 985   \\
                                    &                                    & MEA                                & 82.45$\pm$0.47  & 92.92$\pm$0.38  & 81.36$\pm$0.53  & \textbf{60.34$\pm$3.89}  & 3810  \\
                                    &                                    & AdvMEA                             & 46.40$\pm$6.23  & 46.18$\pm$6.32  & 34.27$\pm$12.78 & 40.85$\pm$8.23  & 150   \\
                                    &                                    & CEGA                               & 81.80$\pm$1.56  & 83.87$\pm$1.15  & 81.64$\pm$1.39  & 49.91$\pm$3.03  & 984   \\
                                    &                                    & DFEAII-Real                        & 82.72$\pm$1.04  & 84.58$\pm$0.58  & 82.42$\pm$0.98  & 48.89$\pm$2.31  & 985   \\
                                    &                                    & Dagger                             & \textbf{84.01$\pm$0.09}  & \textbf{93.44$\pm$0.24}  & \textbf{82.95$\pm$0.07}  & 57.78$\pm$1.89  & 985   \\ \hline

\multirow{24}{*}{\textbf{Physics}}  & \multirow{6}{*}{GCN}               & Rand                               & 50.58$\pm$0.00  & 50.55$\pm$0.00  & 13.44$\pm$0.00  & 26.67$\pm$0.00  & 1724  \\
                                    &                                    & MEA                                & 95.34$\pm$0.11  & \textbf{97.64$\pm$0.21}  & 93.64$\pm$0.15  & \textbf{60.00$\pm$4.80}  & 16072 \\
                                    &                                    & AdvMEA                             & 91.61$\pm$0.33  & 93.19$\pm$0.51  & 88.04$\pm$0.54  & 53.33$\pm$4.80  & 250   \\
                                    &                                    & CEGA                               & 91.92$\pm$0.49  & 93.67$\pm$0.55  & 89.49$\pm$0.55  & 45.93$\pm$2.77  & 1720  \\
                                    &                                    & DFEAII-Real                        & 92.46$\pm$0.50  & 93.94$\pm$0.55  & 89.62$\pm$0.84  & 40.74$\pm$3.78  & 1724  \\
                                    &                                    & Dagger                             & \textbf{95.51$\pm$0.09}  & 97.23$\pm$0.05  & \textbf{93.94$\pm$0.10}  & 59.26$\pm$2.77  & 1724  \\ \cline{2-8}
                                    & \multirow{6}{*}{GAT}               & Rand                               & 50.58$\pm$0.00  & 50.26$\pm$0.00  & 13.44$\pm$0.00  & 25.00$\pm$0.00  & 1724  \\
                                    &                                    & MEA                                & 88.98$\pm$0.81  & 89.94$\pm$0.72  & 85.74$\pm$0.67  & 46.97$\pm$2.14  & 16072 \\
                                    &                                    & AdvMEA                             & 79.46$\pm$0.46  & 80.01$\pm$0.45  & 72.60$\pm$1.46  & 42.42$\pm$4.29  & 250   \\
                                    &                                    & CEGA                               & 88.54$\pm$1.52  & 88.93$\pm$1.22  & 84.99$\pm$1.31  & 50.76$\pm$2.83  & 1720  \\
                                    &                                    & DFEAII-Real                        & 88.41$\pm$1.65  & 88.95$\pm$1.64  & 83.39$\pm$2.234 & 48.48$\pm$7.03  & 1724  \\
                                    &                                    & Dagger                             & \textbf{95.22$\pm$0.05}  & \textbf{94.98$\pm$0.04}  & \textbf{93.55$\pm$0.06}  & \textbf{56.06$\pm$2.14}  & 1724  \\ \cline{2-8}
                                    & \multirow{6}{*}{APPNP}             & Rand                               & 69.76$\pm$0.29  & 70.97$\pm$0.36  & 50.14$\pm$0.16  & 36.60$\pm$3.33  & 1724  \\
                                    &                                    & MEA                                & \textbf{95.54$\pm$0.20}  & \textbf{98.02$\pm$0.14}  & \textbf{93.93$\pm$0.33}  & 59.48$\pm$4.03  & 16072 \\
                                    &                                    & AdvMEA                             & 92.81$\pm$0.46  & 94.93$\pm$0.52  & 89.86$\pm$0.86  & \textbf{60.78$\pm$2.77}  & 250   \\
                                    &                                    & CEGA                               & 70.73$\pm$21.49 & 71.62$\pm$21.77 & 73.10$\pm$15.06 & 42.48$\pm$7.39  & 1720  \\
                                    &                                    & DFEAII-Real                        & 78.25$\pm$3.13  & 78.84$\pm$3.46  & 59.96$\pm$6.02  & 39.87$\pm$6.67  & 1724  \\
                                    &                                    & Dagger                             & 95.46$\pm$0.09  & 97.60$\pm$0.09  & 93.86$\pm$0.12  & 54.90$\pm$1.60  & 1724  \\ \cline{2-8}
                                    & \multirow{6}{*}{GraphSAGE}         & Rand                               & 50.58$\pm$0.00  & 50.55$\pm$0.00  & 13.44$\pm$0.00  & 27.91$\pm$0.00  & 1724  \\
                                    &                                    & MEA                                & 94.71$\pm$0.18  & 96.13$\pm$0.26  & 92.79$\pm$0.25  & 47.29$\pm$2.90  & 16072 \\
                                    &                                    & AdvMEA                             & 91.35$\pm$0.24  & 92.41$\pm$0.18  & 87.83$\pm$0.31  & 49.61$\pm$2.19  & 250   \\
                                    &                                    & CEGA                               & 92.76$\pm$1.49  & 92.81$\pm$1.31  & 91.17$\pm$1.38  & 39.53$\pm$0.00  & 1720  \\
                                    &                                    & DFEAII-Real                        & 93.39$\pm$0.13  & 92.41$\pm$0.18  & 87.83$\pm$0.31  & 49.61$\pm$2.19  & 1724  \\
                                    &                                    & Dagger                             & \textbf{95.42$\pm$0.06}  & \textbf{96.42$\pm$0.05}  & \textbf{93.96$\pm$0.07}  & \textbf{51.94$\pm$2.19}  & 1724 \\ \bottomrule
\end{tabular}}
\label{main_table_app}
\end{table}

\subsection{Balance and Its Impact Factors}
\label{sec_balance}
\noindent\textbf{Balance} To investigate the effect of class balance on attack performance, we construct four balance-level query node sets on Cora, the representative of the most imbalanced graphs. We subsample the full training nodes in the victim graph according to different target distributions. Specifically, let $n_c$ denote the number of training nodes whose victim-predicted label is class $c$, and $|\mathcal{G}_t|$ represents the number of training nodes. The four levels are defined by the following sampling proportions, $p_c^{\text{very}}\!\propto\!\frac{n_c}{|\mathcal{G}_t|}$; $p_c^{\text{imb}}\!\propto\!\sqrt{\frac{n_c}{|\mathcal{G}_t|}}$; $p_c^{\text{slight}}\!\propto\!\left(\frac{n_c}{|\mathcal{G}_t|}\right)^{1/3}$; and $p_c^{\text{balanced}}\!=\!\frac{1}{|C|}$. Each balance level is normalized to sum to one, and the largest-remainder allocation method is used to ensure a fixed query budget across all levels, where each class is first assigned its integer quota of query nodes and the remaining unallocated spots are then sequentially distributed to classes with the largest fractional remainders. The very imbalanced level preserves the heavily skewed distribution of victim predictions on the induced subgraph; imbalanced and slightly imbalanced levels progressively mitigate imbalance issues; and the balanced level enforces strict equality across classes. This design enables a controlled study of class balance effects while maintaining a fixed total query budget.

Contrary to the intuition that perfectly balanced queries should always yield the best surrogate, gradually enforcing balance constraints does not consistently improve attack performance across all victim backbones, sometimes even degrading it. Under high imbalance, the gap between the settings with and without class-balanced sampling and logit adjustment is largest across victim backbones. As the balance level increases, the two curves converge, indicating that rebalancing provides marginal benefit as the query distribution becomes more uniform. Sometimes they might cause a slight performance degradation, one plausible explanation is that excess resampling and calibration enforcements might cause overfitting. 

\begin{figure*}
    \centering
    \begin{subfigure}[b]{0.95\linewidth}
        \centering
        \includegraphics[width=\linewidth]{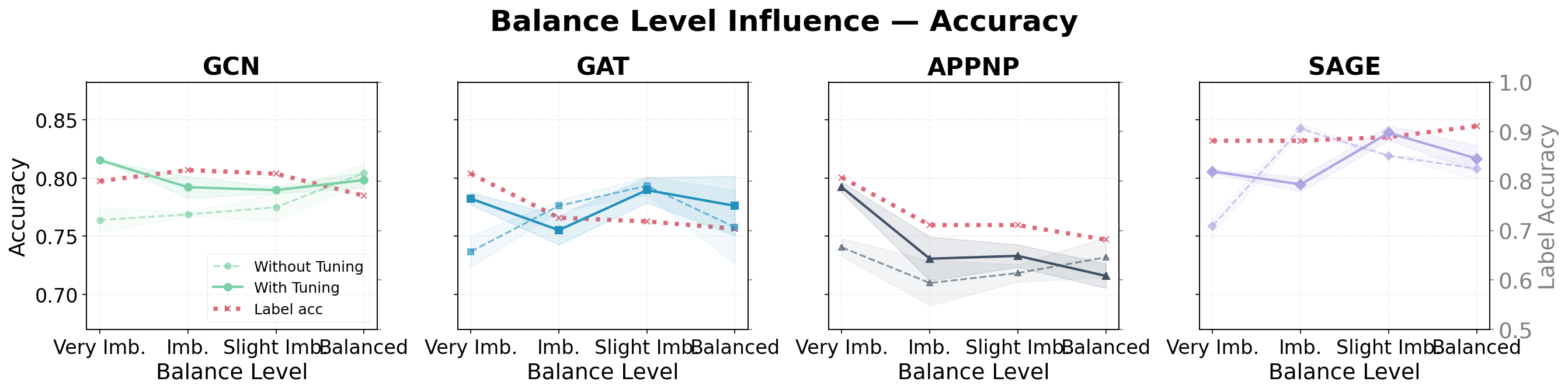}
        \label{fig:placeholder}
    \end{subfigure}

    \begin{subfigure}[b]{0.95\linewidth}
        \centering
        \includegraphics[width=\linewidth]{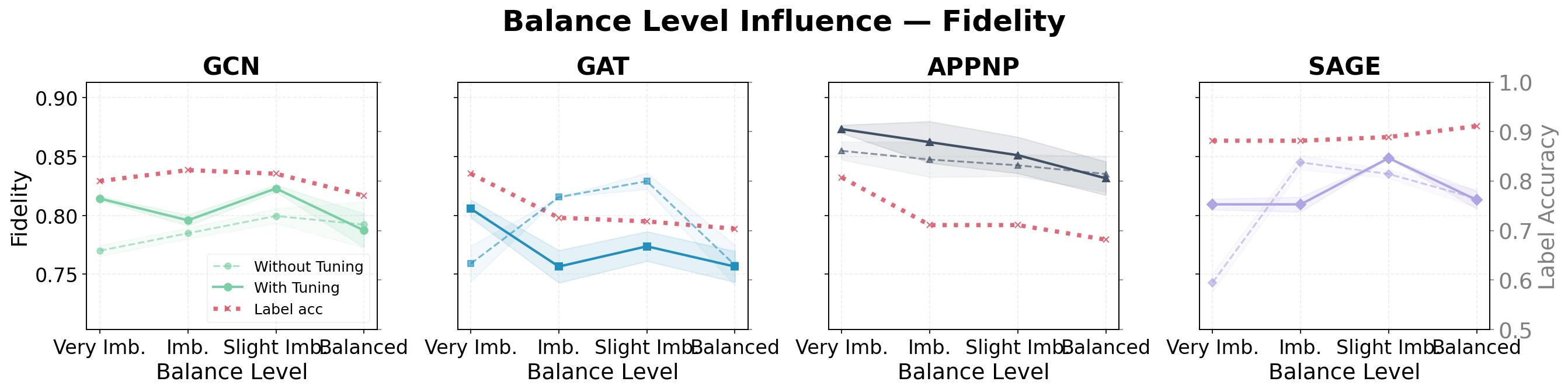}
        \label{fig:placeholder}
    \end{subfigure}

    \begin{subfigure}[b]{0.95\linewidth}
        \centering
        \includegraphics[width=\linewidth]{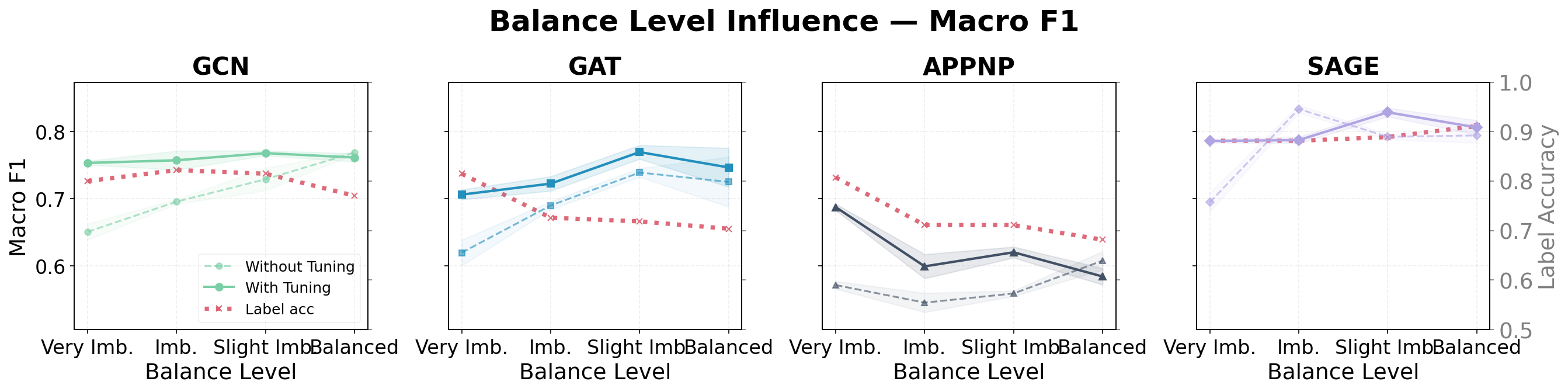}
        \label{fig:placeholder}
    \end{subfigure}

    \begin{subfigure}[b]{0.95\linewidth}
        \centering
        \includegraphics[width=\linewidth]{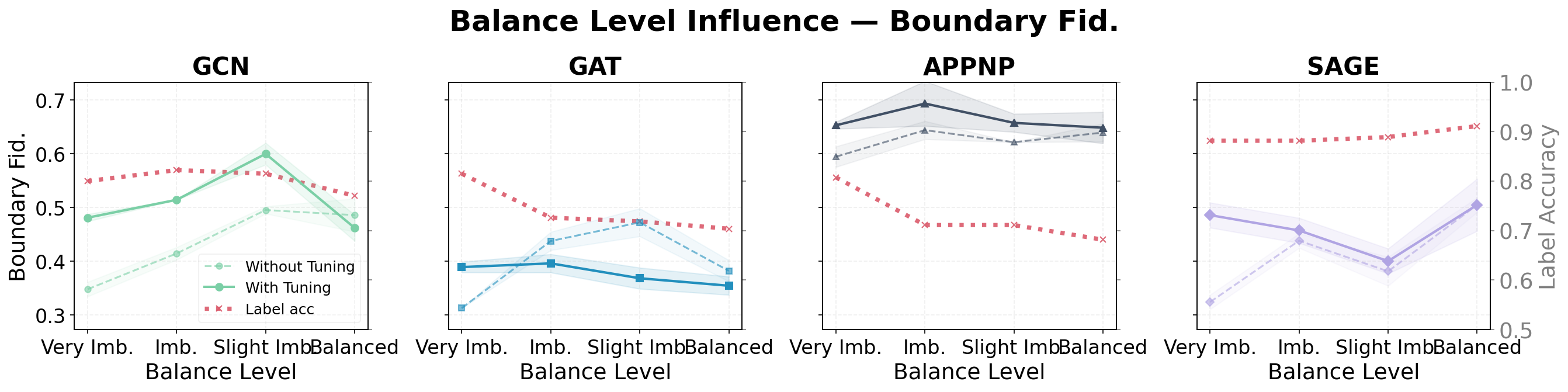}
        \label{fig:placeholder}
    \end{subfigure}

    \begin{subfigure}[b]{0.95\linewidth}
        \centering
        \includegraphics[width=\linewidth]{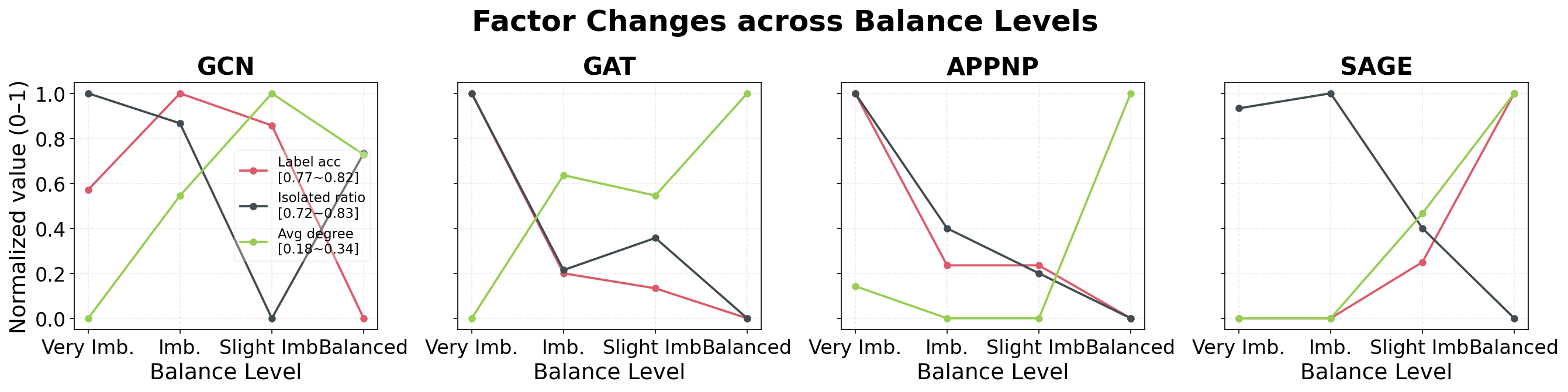}
        \label{fig:placeholder}
    \end{subfigure}
    \caption{Balance Level Sensitivity and Other Factors}
    \label{balance_impact}
\end{figure*}

\noindent\textbf{Impact Factors} We then investigate this finding with three factors interlinked with balance, namely label accuracy, isolated ratio and average degree in Figure \ref{balance_impact}. As the balance level increases, these three factors exhibit victim-dependent and non-monotonic trends across balance levels, suggesting that the relationship between query balance and data quality is more nuanced than a simple trade-off. 

Label accuracy shows inconsistent trends across victim backbones: it first increases then decreases for GCN, decreases then stably increases for GAT and APPNP, and steadily increases for GraphSAGE as the balance level rises. This suggests that the label quality effect of more balanced node selection is highly dependent on the victim's prediction behaviors. For some backbones, selectively including minority-class nodes can eventually improve the overall label accuracy of the query set, while for others it introduces noise.

Isolated ratio is similarly non-monotonic: it decreases then rises for GCN, decreases then partially recovers for GAT, decreases monotonically for APPNP, and first increases then decreases for GraphSAGE. This indicates a complex interplay between query class balance and the structural sparsity of local subgraphs, where enforcing class balance does not guarantee structural integrity in the query set.

Average degree also exhibits victim-specific patterns: it rises then falls for GCN, rises with a slight dip then rises again for GAT, slightly decreases then sharply increases for APPNP, and remains stable before rising for GraphSAGE, which showcases that the structural richness of the query set does not vary predictably with query balance level.

\subsection{Ablation Studies}
\begin{table}[htbp]
\caption{\centering Ablation Studies across Graph Datasets and GNN Backbones}
\centering
\resizebox{\linewidth}{!}{
\begin{tabular}{c|c|c|cccc}
\toprule
\multirow{2}{*}{\textbf{Dataset}}   & \multirow{2}{*}{\textbf{Backbone}} & \multirow{2}{*}{\textbf{Ablation}} & \multicolumn{4}{c}{\textbf{Metrics}}                              \\ \cline{4-7}
                                    &                                    &                                    & ACC            & FID            & Macro F1              & Boundary Fid.          \\ \hline
\multirow{16}{*}{\textbf{PubMed}}   & \multirow{4}{*}{GCN}               & No manifold                        & 83.67$\pm$0.08 & 91.48$\pm$0.23 & 82.50$\pm$0.18 & 52.78$\pm$0.21 \\
                                    &                                    & No LA                              & \textbf{84.18$\pm$0.11} & \textbf{91.77$\pm$0.42} & \textbf{83.12$\pm$0.18} & 53.65$\pm$1.61 \\
                                    &                                    & No Tuning                         & 83.54$\pm$0.71 & 91.21$\pm$0.48 & 82.11$\pm$0.98 & 52.78$\pm$1.09 \\
                                    &                                    & Dagger                             & 83.81$\pm$0.06 & 91.72$\pm$0.17 & 82.80$\pm$0.14 & \textbf{54.53$\pm$0.55} \\
                                    \cline{2-7}
                                    & \multirow{4}{*}{GAT}               & No manifold                        & \textbf{84.08$\pm$0.43} & 92.11$\pm$0.16 & \textbf{82.83$\pm$0.50} & 59.07$\pm$0.42 \\
                                    &                                    & No LA                              & 83.89$\pm$0.06 & 92.26$\pm$0,17 & 82.75$\pm$0.11 & 57.34$\pm$1.49 \\
                                    &                                    & No Tuning                         & 83.81$\pm$0.19 & \textbf{92.33$\pm$0.53} & 82.60$\pm$0.17 & \textbf{60.45$\pm$2.48} \\
                                    &                                    & Dagger                             & 83.54$\pm$0.32 & 92.29$\pm$0.27 & 82.41$\pm$0.33 & 56.65$\pm$1.29 \\
                                    \cline{2-7}
                                    & \multirow{4}{*}{APPNP}             & No manifold                        & \textbf{83.54$\pm$0.23} & \textbf{94.46$\pm$0.49} & \textbf{82.56$\pm$0.17} & \textbf{63.97$\pm$1.75} \\
                                    &                                    & No LA                              & 83.42$\pm$0.17 & 94.34$\pm$0.12 & 82.39$\pm$0.17 & 63.43$\pm$1.01 \\
                                    &                                    & No Tuning                         & 83.35$\pm$0.25 & 94.15$\pm$0.53 & 82.31$\pm$0.18 & 63.16$\pm$1.72 \\
                                    &                                    & Dagger                             & 83.27$\pm$0.15 & 93.68$\pm$0.13 & 82.29$\pm$0.14 & 62.48$\pm$1.16 \\
                                    \cline{2-7}
                                    & \multirow{4}{*}{GraphSAGE}         & No manifold                        & 84.30$\pm$0.20 & 93.36$\pm$0.47 & 83.15$\pm$0.12 & 59.32$\pm$2.85 \\
                                    &                                    & No LA                              & 84.31$\pm$0.38 & 93.31$\pm$0.37 & 83.25$\pm$0.31 & 57.78$\pm$1.47 \\
                                    &                                    & No Tuning                         & \textbf{84.48$\pm$0.35} & \textbf{93.54$\pm$0.67} & \textbf{83.33$\pm$0.27} & \textbf{59.66$\pm$3.75} \\
                                    &                                    & Dagger                             & 84.01$\pm$0.09 & 93.44$\pm$0.24 & 82.95$\pm$0.07 & 57.78$\pm$1.89 \\ \hline
\multirow{16}{*}{\textbf{Physics}}  & \multirow{4}{*}{GCN}               & No manifold                        & 94.33$\pm$0.05 & 95.46$\pm$0.09 & 92.21$\pm$0.11 & 42.96$\pm$1.05 \\
                                    &                                    & No LA                              & 95.04$\pm$0.23 & 96.44$\pm$0.15 & 93.40$\pm$0.34 & 49.63$\pm$2.10 \\
                                    &                                    & No Tuning                         & 94.86$\pm$0.18 & 96.53$\pm$0.12 & 93.26$\pm$0.26 & 47.41$\pm$4.57 \\
                                    &                                    & Dagger                             & \textbf{95.51$\pm$0.09} & \textbf{97.23$\pm$0.05} & \textbf{93.94$\pm$0.10} & \textbf{59.26$\pm$2.77} \\ \cline{2-7}
                                    & \multirow{4}{*}{GAT}               & No manifold                        & 93.68$\pm$0.21 & 93.54$\pm$0.12 & 91.58$\pm$0.30 & 48.48$\pm$2.14 \\
                                    &                                    & No LA                              & 94.80$\pm$0.18 & 94.36$\pm$0.21 & 93.08$\pm$0.18 & 52.27$\pm$1.86 \\
                                    &                                    & No Tuning                         & 94.96$\pm$0.05 & 94.53$\pm$0.01 & 93.30$\pm$0.09 & 51.52$\pm$2.14 \\
                                    &                                    & Dagger                             & \textbf{95.22$\pm$0.05} & \textbf{94.98$\pm$0.04} & \textbf{93.55$\pm$0.06} & \textbf{56.06$\pm$2.14} \\ \cline{2-7}
                                    & \multirow{4}{*}{APPNP}             & No manifold                        & 94.87$\pm$0.00 & 96.68$\pm$0.04 & 92.92$\pm$0.03 & 54.90$\pm$0.00 \\
                                    &                                    & No LA                              & 95.70$\pm$0.12 & \textbf{97.89$\pm$0.11} & 94.28$\pm$0.18 & \textbf{58.17$\pm$0.92} \\
                                    &                                    & No Tuning                         & \textbf{95.77$\pm$0.06} & 97.82$\pm$0.03 & \textbf{94.39$\pm$0.04} & \textbf{58.17$\pm$2.45} \\
                                    &                                    & Dagger                             & 95.46$\pm$0.09 & 97.60$\pm$0.09 & 93.86$\pm$0.12 & 54.90$\pm$1.60 \\ \cline{2-7}
                                    & \multirow{4}{*}{GraphSAGE}         & No manifold                        & 94.89$\pm$0.08 & 95.65$\pm$0.10 & 92.74$\pm$0.11 & 42.96$\pm$1.05 \\
                                    &                                    & No LA                              & 95.04$\pm$0.23 & \textbf{96.44$\pm$0.15} & 93.40$\pm$0.34 & 49.63$\pm$2.10 \\
                                    &                                    & No Tuning                         & 95.08$\pm$0.05 & 95.74$\pm$0.05 & 93.34$\pm$0.08 & 44.96$\pm$1.10 \\
                                    &                                    & Dagger                             & \textbf{95.42$\pm$0.06} & 96.42$\pm$0.05 & \textbf{93.96$\pm$0.07} & \textbf{51.94$\pm$2.19}
                                    \\ \bottomrule

\end{tabular}}
\label{ablation_2}
\end{table}

\end{document}